\documentclass[%
 reprint,
superscriptaddress,
nofootinbib,
 amsmath,amssymb,
 aps,
]{revtex4-2}

\usepackage{graphicx}% Include figure files
\usepackage{dcolumn}% Align table columns on decimal point
\usepackage{hyperref}
\usepackage{fontawesome5}
\usepackage{bm}
\usepackage{orcidlink}
\usepackage{CJKutf8}

\usepackage{xcolor}
\usepackage{color}
\usepackage{hyperref}

\newcommand{\fionaedit}[1]{{\textcolor{magenta}{{#1}}}}

\usepackage[normalem]{ulem}

\newcommand{\guanmingedit}[1]{{\textcolor{blue}{{#1}}}}

\usepackage[normalem]{ulem}

\allowdisplaybreaks

\begin{document}

\preprint{APS/123-QED}

\title{Constraining Local Primordial Non-Gaussianity with Cross Correlations Alone}% Force line breaks with \\
% \thanks{A footnote to the article title \frank{what does this footprint mean}}

\author{\begin{CJK*}{UTF8}{gbsn} Guanming Liang (梁冠铭)\end{CJK*} \orcidlink{0000-0002-2581-2406}}
\email{Correspondence: gl570@cam.ac.uk}
\affiliation{%
 DAMTP, Centre for Mathematical Sciences, University of Cambridge, Wilberforce Road,
Cambridge CB3 0WA, UK
}%
\affiliation{Kavli Institute for Cosmology Cambridge, Madingley Road, Cambridge CB3 0HA, UK}

\author{Fiona McCarthy}
\affiliation{Institute of Astronomy, University of Cambridge, Madingley Road, Cambridge CB3 0HA, UK}
\affiliation{Kavli Institute for Cosmology Cambridge, Madingley Road, Cambridge CB3 0HA, UK}
\affiliation{%
 DAMTP, Centre for Mathematical Sciences, University of Cambridge, Wilberforce Road,
Cambridge CB3 0WA, UK
}%
\author{Blake D.~Sherwin}
\affiliation{%
 DAMTP, Centre for Mathematical Sciences, University of Cambridge, Wilberforce Road,
Cambridge CB3 0WA, UK
}%
\affiliation{Kavli Institute for Cosmology Cambridge, Madingley Road, Cambridge CB3 0HA, UK}

\author{Frank J. Qu \,\orcidlink{0000-0001-7805-1068}}
 \affiliation{Kavli Institute for Particle Astrophysics and Cosmology, Stanford University, 452 Lomita Mall, Stanford, CA, 94305, USA}
 \affiliation{Department of Physics, Stanford University, 382 Via Pueblo Mall, Stanford, CA, 94305, USA} \affiliation{NASA Hubble Fellowship Program, Einstein Fellow}

\date{\today}

\begin{abstract}
A key target of ongoing and upcoming galaxy surveys is to measure local primordial non-Gaussianity via the scale-dependent bias. However, measurements of large-scale bias via galaxy clustering can be challenging due to observational systematic errors, which motivates alternative approaches to constraining the relevant parameter, $f_{\rm NL}^{\rm loc}$ (hereafter $f_{\rm NL}$). In this paper, we forecast to what extent $f_{\rm NL}$ can be constrained from cross-correlations of galaxies with external tracers, with the large-scale galaxy auto-spectrum discarded. In particular, we focus on correlations with kSZ velocity reconstruction as well as with CMB gravitational lensing. We consider both the SPHEREx and LSST surveys in combination with a Simons-Observatory-like survey, and perform a detailed exploration of the dependence of our results on analysis choices. We find that competitive kSZ cross-correlation-derived constraints on $f_{\rm NL}$ can be obtained: with LSST (SPHEREx), we forecast errorbars of $\sigma(f_{\rm NL})\sim 0.9-1.9$ ($1.0–1.35$), where the range reflects the uncertain amplitude of the kSZ signal. The combination of lensing and kSZ cross-correlations is even more promising, with $\sigma(f_{\rm NL})\sim 1.35$ ($\sim 1.15$) obtainable for LSST (SPHEREx) even for a low kSZ amplitude. Cross-correlations therefore offer an independent path, with different systematics to galaxy clustering, towards the $\sigma(f_{\rm NL})\sim 1$ threshold between single-field and multi-field inflation.
\end{abstract}

\maketitle

%\tableofcontents

\section{Introduction}
An initial accelerated expansion phase of the Universe is a compelling mechanism that solves several problems of hot big bang cosmology and provides a mechanism to explain structure in the Universe by stretching initially microscopic quantum fluctuations to produce the primordial perturbations. However, we do not have many constraints on the physical nature of inflation, such as the precise physical mechanism and the sourcing field(s). The simplest picture consistent with current observations is inflation driven by a single scalar field slowly rolling down a potential. However, it is plausible that several fields may have been active. Constraints on the (non-)Gaussianity of the initial conditions of the Universe are valuable pathways to distinguishing between different inflationary mechanisms, as different mechanisms  lead to different amounts and types of primordial non-Gaussianity (PNG). In particular, inflation sourced by multiple interacting scalar fields leads to local-type non-Gaussianity, which is of the form
\begin{equation}
\Phi(\mathbf{x})=\Phi_G(\mathbf{x})+f_{\rm NL}^{\rm loc}(\Phi^2_G(\mathbf{x})-\langle\Phi^2_G(\mathbf{x})\rangle),
\end{equation}
where $\Phi(\mathbf{x})$ is the primordial potential at location $\mathbf{x}$, $\Phi_G(\mathbf{x})$ is a Gaussian field, angular brackets indicate the expectation value, and the local-type non-Gaussianity is quantified by $f_{\rm NL}^{\rm loc}$ (henceforward $f_{\rm NL}$).

{A measurement of nonzero $f_{\rm NL}$ would definitively rule out the single-field slow-roll paradigm \cite{Maldacena:2002vr, Creminelli:2004yq, Pajer:2013ana,dePutter:2015vga}, and ruling out $f_{\rm NL}\gtrsim \mathcal{O}(1)$ would disfavor multi-field inflation}. Thus, the cosmology community is targeting a measurement with errorbar~$\sigma(f_{\rm NL})\lesssim 1$ as the critical threshold for near-future PNG measurements. The leading constraint to date~\cite{Planck:2019kim, Jung:2025nss} from the {cosmic microwave background (CMB),~$f_{\rm NL}=-0.1\pm 5.0$}, was extracted from the ``squeezed" bispectrum of the \textit{Planck} PR4 CMB temperature and E-mode polarization maps, where one long-wavelength (large-scale) mode modulates two short-wavelength (small-scale) modes.\footnote{More recently, Ref.~\cite{Chudaykin:2025vdh} used the power spectrum and bispectrum extracted from DESI DR1 \cite{DESI:2025fxa} in combination with Legacy BOSS DR12 \cite{BOSS:2016wmc} {and \textit{Planck} CMB \cite{Planck:2019kim} data} to constrain $f_{\rm NL}=0.0\pm 4.1$, stronger than the leading CMB-only \textit{Planck} constraint, which dominated this constraint.} Once the CMB polarization and small-scale temperature modes are mostly exhausted, improvements on $\sigma(f_{\rm NL})$ from the CMB will be marginal due to cosmic variance; for the ongoing Simons Observatory (SO) \cite{SimonsObservatory:2018koc,SimonsObservatory:2025wwn}, which will map the smaller-scale (up to~$\sim 1 ^\prime$) CMB modes at higher precision than ever before on wide sky area, the CMB primary bispectrum alone is forecasted to give $\sigma(f_{\rm NL})\sim 3$~\cite{SimonsObservatory:2018koc}.

% \textit{Scale-dependent bias on galaxies.}
Another promising local PNG probe, expected to yield tighter constraints than the CMB in the near future, is galaxy overdensity {$\delta_g$}. This is defined as the {position-dependent fractional deviation} from the average number density of galaxies. It is a biased tracer of the underlying matter density contrast ($\delta_m$) with $\delta_g=b_g\delta_m$ at linear level, {where $b_g$ (which could be a function of $z$) defines the galaxy bias}. In a Gaussian universe, linear galaxy bias is independent of scale. In the presence of local PNG, galaxy bias receives a correction that scales as $\Delta b_g(k) \propto f_{\rm NL}/k^{2}$ \cite{Dalal:2007cu,2008JCAP...08..031S,2014PhRvL.113v1301L,2014PhRvD..89b3511G,2014MNRAS.441L..16G,2022MNRAS.514.3396M} (where $k$ is the wavenumber and indicates the scale), making large scales of the galaxy distribution the most sensitive to local PNG. 

However, there are various observational systematics that make it challenging to interpret galaxy clustering measurements at large scales, such as the following examples. Spurious stars can be mistaken as galaxies and result in over-counting of the galaxy number density. Non-uniform sensitivity across the sky due to, for example, galactic dust extinction gives rise to a similar {miscount}: interstellar/intergalactic dust can cause colour reddening of distant (high-redshift) galaxies in certain directions and an underestimation of their number density. There are further terrestrial systematics like atmospheric turbulence and sky brightness. These systematics are dominant at large scales. As shown in Figure~11 of Ref.~\cite{Rezaie:2023lvi}, the galaxy large-scale angular spectrum is highly dependent upon the exact imaging mitigation technique. Such systematics are hence challenging to disentangle from the cosmological, large-scale galaxy clustering signal where $f_{\rm NL}$ happens to leave the most detectable imprint. 

Although much progress has been made on mitigating these autospectrum systematics in modern analyses, alternative approaches with different systematics are highly valuable, especially in the case that a detection is made and requires confirmation from alternative probes. We propose future analyses exploit the advantage that cross correlations between independent observables do not suffer from additive systematic bias. {Previous combined-probes forecasts \cite{Schmittfull:2017ffw,Munchmeyer:2018eey} have mostly considered how different tracers of the same underlying large-scale structure (LSS) share the same realization of~$\delta_m$ and thus cancel out its stochastic variance in a process called cosmic variance cancellation (CVC)~\cite{Seljak:2008xr}, allowing for \textit{improved} constraints compared to a galaxies-alone case, whereas we emphasize here that being less prone to systematics (perhaps at the cost of some signal-to-noise) is another crucial advantage of cross correlations. In this work, we explore their constraining power without the galaxy auto power spectrum {at the largest scales}.

CMB secondary effects come into focus in such cross correlations. These arise from interactions between CMB photons and the LSS (e.g. via gravity or electromagnetic scattering) as the photons travel from the surface of last scattering to us today, {thus emerging} as powerful tracers of the matter distribution. {The} cross correlation of galaxies with {highly correlated} CMB secondaries will be powerful for advancing future $f_{\rm NL}$ measurements. One notable CMB secondary particularly suited for this purpose is the kinetic Sunyaev--Zel'dovich (kSZ) effect, arising from Thomson scattering of CMB photons off free electrons in the ionized intergalactic medium \cite{Sunyaev:1980nv,2012PhRvL.109d1101H}. Due to the bulk peculiar {velocity} $\mathbf{v}$ of the ionized gas relative to the CMB rest frame, electrons experience a Doppler shift and observe a local CMB dipole \cite{Terrana:2016xvc}. The dominant contribution to the resulting observed temperature anisotropy from this kSZ signal, $T_{\rm kSZ}$, is proportional to the radial velocity ${v_r}\equiv\mathbf{v}\cdot\hat{n}$ weighted by the electron density $n_e$, integrated {over comoving distance $\chi$} along the line of sight (LOS, defined by $\hat n$): $T_{\rm kSZ}\sim\int d\chi n_ev_r$. The kSZ contribution is subdominant to the lensed primary CMB on large-to-intermediate scales but becomes dominant on small scales as shown in Fig.~\ref{fig:act_v_so_cltt} (angular multipole $l\gtrsim 4000$ or angular resolutions $\theta\sim 1^\prime$). With low-noise high-angular-resolution maps of the CMB, e.g. from SO, in combination with high number-density galaxy overdensity at small scales from the current generation of ongoing surveys such as Vera Rubin Observatory's Legacy Survey of Space and Time (LSST)\cite{LSSTScience:2009jmu}, SPHEREx~\cite{SPHEREx:2014bgr, Bock:2025ijl}, DESI~\cite{DESI:2022xcl}, Euclid~\cite{Euclid:2024yrr}, and Roman~\cite{NANCY:2023mzs}, we can create with high fidelity a reconstruction $\hat{v}$ of the large-scale radial velocity field $v_r$ with a quadratic estimator~\cite{Deutsch:2017ybc,Smith:2018bpn}. It was first shown in Ref.~\cite{Munchmeyer:2018eey} that the kSZ-reconstructed velocity gives a highly-correlated tracer to cross-correlate with $\delta_g$ and exploit CVC for constraining~$f_{\mathrm{NL}}$. The best-reconstructed modes in kSZ velocity reconstruction are at the largest scales, which is why it is so useful in $f_{\rm NL}$ analyses.

\begin{figure}[h]
    \centering
    \includegraphics[width=\linewidth]{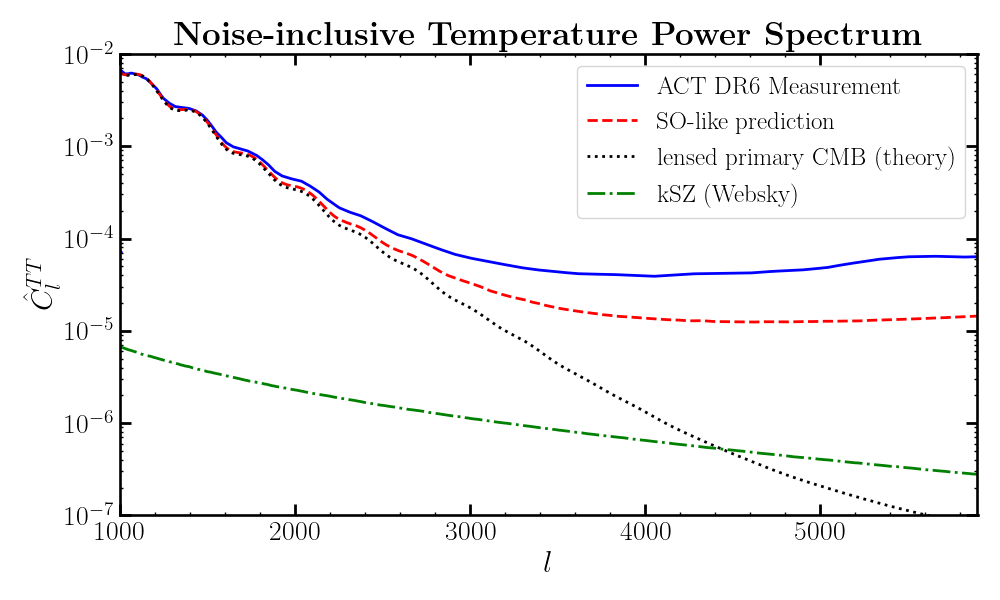}
    \caption{ACT compared to SO-like CMB temperature signal and noise curves, consisting of the lensed primary CMB, residual foregrounds after cleaning, and instrumental noise. The primary CMB is computed with CAMB. The ACT curve comes from a power spectrum measurement of the ACT DR6 NILC blackbody temperature map~\cite{2024PhRvD.109f3530C,2025JCAP...11..061N}. The SO-like prediction adds the forecasted, post-component-separated noise for SO to the primary CMB. The kSZ prediction comes from the measurement of power spectrum of the kSZ component of the Websky simulations~\cite{2020JCAP...10..012S}. We see that the kSZ contribution to the power spectrum is expected to dominate over the lensed primary CMB at $l\sim 4400$.} 
    \label{fig:act_v_so_cltt}
\end{figure}

% Heretofore, all concrete kSZ {velocity} detections have relied on cross correlations with other better measured observables, such as with galaxies in $gv$ \cite{McCarthy:2024nik,2025PhRvL.134o1003L,Hotinli:2025tul, Lai:2025qdw,McCarthy:2025brx,EmbilVillagra:2026fnh,Chaussidon:2026vmn,Hotinli:2026ruo}, but high-confidence detections of the kSZ velocity-velocity autospectrum $vv$ are expected in the near future, with recent measurements already at $2.1\sigma$ \cite{McCarthy:2025brx}, $3.1\sigma$ \cite{Chaussidon:2026vmn}, {and $8.1\sigma$ \cite{Hotinli:2026ruo}}.

Until recently, all concrete kSZ {velocity} detections have relied on cross correlations with other better measured observables, such as with galaxies in $gv$ \cite{McCarthy:2024nik,2025PhRvL.134o1003L,Hotinli:2025tul, Lai:2025qdw,McCarthy:2025brx,EmbilVillagra:2026fnh,Chaussidon:2026vmn,Hotinli:2026ruo}. The kSZ velocity-velocity autospectrum $vv$ has now been measured at $2.1\sigma$ \cite{McCarthy:2025brx}, $3.1\sigma$ \cite{Chaussidon:2026vmn}, {and $8.1\sigma$ \cite{Hotinli:2026ruo}}.

Meanwhile, weak gravitational lensing of the CMB~\cite{Lewis:2006fu} is directly sensitive to the total matter distribution, mapping the integrated matter distribution of the Universe. As CMB photons travel from the last scattering surface, their trajectories are deflected by gravity from intervening matter structures. This process physically remaps the 
unlensed primary CMB temperature field in direction $\hat n$, {$T(\hat{n})$, to the lensed temperature field} $\tilde{T}(\hat{n})=T(\hat{n} + {\vec{\alpha}})$, where the deflection angle $\vec{\alpha}=\vec{\nabla}\phi$ is the gradient of the lensing potential $\phi$, which is itself a line-of-sight projection of the gravitational potential $\Psi$. Since $\Psi$ directly responds to $\delta_m$ via the Poisson equation, $\nabla^2\Psi=3\Omega_{m,0}H_0^2\delta_m/2a$ (where $\Omega_{m,0}$ is the matter density today, $H_0$ is the Hubble parameter today, and $a$ is the scale factor), CMB lensing is an unbiased tracer of matter overdensity and can be similarly exploited for CVC~\cite{Schmittfull:2017ffw}. It is commonly studied through its convergence $\kappa\equiv -\vec{\nabla}\cdot\vec{\alpha}/2$. Our focus is how accurately we will measure $f_{\rm NL}$ if it is constrained at large scales by only cross correlations between these three tracers ~$\delta_g$,~$\kappa$, and $v_r$ (we still include the autospectra of $v_r$ and $\kappa$, which do not depend on $f_{\rm NL}$, although we also explore constraints which drop these as well). In this paper, we calculate explicitly the errorbar on $f_{\rm NL}$ for an SO-like+LSST-like set up as well as an SO-like+SPHEREx-like setup.
% \textit{Organization.} 

The paper is organized as follows. In Sec.~\ref{basics}, we briefly review 2D redshift tomography, kSZ velocity reconstruction, and the kSZ optical depth degeneracy. Then in Sec.~\ref{forecasts}, we show detailed $gv$ and $g\kappa$ cross correlation forecasts for LSST and SPHEREx, and explore the improvement when simultaneously cross-correlating both tracers  with galaxies. We} summarize our results in Sec.~\ref{result}, and conclude in Sec.~\ref{concl}.

\section{Basic Setup}\label{basics}
{Galaxy bias becomes scale-dependent in the presence of $f_{\rm NL}$} according to
\begin{align}\label{scale-dep_bias}
    b_g(z)\rightarrow b_g^{\rm NL}(z)=b_g(z)+\frac{f_{\rm NL}}{k^2}\frac{(b_{g}(z)-1)}{\alpha(k,z)},
\end{align}
where $\alpha(k,z)=T(k)D(z)/(3\Omega_{m,0}H_0^2\delta_c)$. We assume universality of the halo mass function{, where the local PNG bias parameter $b_\Phi = 2\delta_c(b_g - p)$ with $p=1$}. Here, $\delta_c$ is the critical threshold above which overdensities collapse; $D(z)$ is the growth function normalized such that $D(z)=1/(1+z)$ during matter domination; and $T(k)$ is the transfer function, which is $\simeq 1$ at the large scales we are concerned with. We take  $\delta_c=1.686$, appropriate for Einstein--de~Sitter~\cite{1972ApJ...176....1G}. Thus, galaxy overdensity is related to the matter overdensity via $\delta_g = b_g^{\rm NL}\delta_m$. 
The scale-dependent bias is imprinted in both the galaxy autospectrum and its cross correlation with other correlated tracers, where it constrains only the product $f_{\rm NL}b_\Phi$~\cite{Barreira:2020ekm, Barreira:2022sey}, so our $\sigma(f_{\rm NL})$ forecasts assume perfect knowledge of $b_\Phi$. For a sample with another $p$, they scale roughly as $(b_g-1)/(b_g-p)$. Assuming $p=1$ when it is $0.55$ biases $f_{\rm NL}$ by $0.5$--$1\sigma$~\cite{Barreira:2020ekm}, and $p\simeq 1.4$--$1.6$ is preferred for quasars~\cite{Fondi:2026ilz,DESI:2023duv}. 

{In our cross correlation forecasts for LSST \cite{LSSTScience:2009jmu} and SPHEREx~\cite{SPHEREx:2014bgr}, the physical observables are the projected galaxy overdensity $\delta_g$, kSZ velocity (i.e., velocity reconstructed from the kSZ signal) $v_{\rm kSZ}$, and CMB lensing convergence $\kappa$. These are two-dimensional projections of the underlying three-dimensional matter field. We present our model for their cross-correlation power spectra below. We refer to $\delta_g$ as $g$ and $v_{\mathrm{kSZ}}$ as $v$ for brevity.} {Throughout, all cosmological calculations are performed with \texttt{CAMB} \cite{Lewis:1999bs}, using the best-fit $\Lambda$CDM cosmological parameters of \textit{Planck} 2018~\cite{Planck:2018vyg}, which are the physical baryon density $\Omega_bh^2=0.0224$, physical CDM density $\Omega_ch^2=0.12$, angular scale of the sound horizon at recombination $100\theta_{\rm MC}=1.0411$, the reionization optical depth $\tau=0.054$, primordial scalar power spectrum amplitude $A_s=2.1\times 10^{-9}$ at the pivot scale $k=0.05$~Mpc$^{-1}$, and the scalar spectral index $n_s=0.965$.\footnote{Note that more recently the ACT DR6+\textit{Planck} analysis~\cite{AtacamaCosmologyTelescope:2025blo} constrained these tighter than \textit{Planck} alone.}}

\subsection{Tomographic Projection}\label{tomography}

For 2D redshift tomography, we consider galaxy surveys with true redshift distribution given by survey-dependent $dn/dz$, binned into $N$ two-dimensional redshift bins labeled by $i${, which are} defined by top-hat bins in photo-redshift (photo-$z$).  We project the {redshift-dependent matter overdensity field $\delta_m(\chi,\hat{n})$} into each redshift bin with projection kernels $W_X$ to calculate the relevant angular power spectra between projected quantities $X, Y\in\{g_i(\hat n), v_i(\hat n), \kappa(\hat n)\}$:%\fiona{in the equation below $v$ is superscript but you described it here with subscript}
\begin{align}
    \kappa(\hat n)&=\int_0^{\chi_{\rm dec}} d\chi W_{\kappa}(\chi)\delta_m(\chi,\hat{n});\\
    g_i(\hat n)&=\int_0^{\chi_{\rm dec}} d\chi W_g^i(k,\chi)\delta_m(\chi,\hat{n});\\
    v_i(\hat n)&=\int_0^{\chi_{\rm dec}} d\chi W_{\rm kSZ}^i(\chi)v_r(\chi,\hat n),
\end{align}
where the upper integration limit $\chi_{\mathrm{dec}}$ indicates the comoving distance to the CMB, and {$v_r(\chi,\hat n)$} can be related at linear level to $\delta_m$ by the continuity equation according to
\begin{align}\label{vcont}
    {v_r(\chi,\hat n)}=\int \frac{d^3k}{(2\pi)^3}i\frac{faH\mu(\hat k)}{k}e^{ik\mu\chi}{\delta_m(\vec{k},\chi)},
\end{align}
where $f$ is the logarithmic growth rate defined as $f(a) = d\ln D(a)/d\ln a$, $\mu = \hat{k}\cdot\hat{n}$ is the cosine of the angle between the unit Fourier vector $\hat{k}$ and the LOS $\hat{n}$, $k$ is the magnitude of $\vec{k}$, and {$\delta_m(\vec{k},\chi) = \int d^3x\, e^{-i\vec{k}\cdot\vec{x}}\, \delta_m(\vec{x},\chi)$} is the Fourier transform of the matter overdensity at the time corresponding to comoving distance $\chi$. The kernels $W_X$ are given by
\begin{align}
    &W_g^{i}(k, \chi) = \frac{b_{g}^{\rm NL}}{\bar{n}_i}\frac{dz}{d\chi}\frac{dn_i}{dz};\\
    &W_{\rm kSZ}^i(\chi) =\frac{b_{v,i}}{\bar{n}_i}\frac{dz}{d\chi}\frac{dn_i}{dz}; \label{kSZ_kernel}\\
    &W_{\kappa}(\chi) = \frac{3H_0^2\Omega_{m,0}}{2a}\frac{\chi_{\rm dec}-\chi}{\chi_{\rm dec}}\chi,
\end{align}
where $b_g^{\rm NL}$ is the galaxy bias defined in Eq.~\eqref{scale-dep_bias}, $\frac{dz}{d\chi}=H(z)$,  $\frac{dn_i}{dz}$ gives the true redshift distribution in redshift bin $i$, and the ``kSZ velocity biases'' $b_{v,i}$ are scale-independent amplitudes sourced by any mismodelling of the small-scale galaxy-electron power spectrum used to estimate $v$ (see Sec.~\ref{ksz recon}). 
For the top-hat bins in photo-$z$ that we work with, $dn_i/dz$ is the overall $dn/dz$ multiplied by the Gaussian probability that a galaxy at true redshift $z$ is measured to be in bin $i$, due to redshift measurement uncertainty $\sigma(z)$ (the upper and lower integration limits come from the top-hat bin in photo-$z$ $z_p$) \cite{Ma:2005rc}:
\begin{equation}\label{eq_dnidz}
    \frac{dn_i}{dz} = \frac{dn}{dz}\int_{z_i}^{z_{i+1}}\guanmingedit{}\frac{1}{\sqrt{2\pi}\sigma(z)}\exp\left(-\frac{(z_p-z)^2}{2\sigma(z)^2}\right)dz_p,
\end{equation}
where $z_i$ and $z_{i+1}$ indicate the lower and upper boundaries respectively of redshift bin $i$.  

Starting with the case where the projected fields are $X, Y\in\{g_i,\kappa\}$, the angular cross-power-spectrum $C_l^{XY}$ is given by
\begin{align}\label{proj}
    C_l^{XY} &= \frac{2}{\pi}\int_{0}^{k_{\rm max}} dk k^2 \int_{0}^{\chi_{\rm dec}} d\chi W_X(k, \chi) j_{l}(k\chi)\nonumber\\
    &\int_{0}^{\chi_{\rm dec}} d \chi^\prime  W_Y(k, \chi^\prime)j_l(k\chi^\prime)P(k,z,z'),
\end{align}
where the three-dimensional (unequal-time) power spectrum $P(k,z,z^\prime)$ is defined through $\langle \delta_m(k,\chi)\delta_m^\ast (k',\chi')\rangle=(2\pi)^3\delta^{(3)}(k-k')P(k, z,z')$, and $j_l(x)$ are the spherical Bessel functions of degree $l$. We model the unequal-time power spectrum as $P(k, z,z')=\mathcal{D}(z)\mathcal{D}(z')P_m(k)$, where $\mathcal{D}(z)$ is the growth factor normalized such that~$\mathcal{D}(z=0)=1$ (note the different normalization of the growth factor $\mathcal{D}(z)$ here from $D(z)$ in Eq.~\eqref{scale-dep_bias}) and $P_{m}(k)$ is the linear matter power spectrum today. The upper limit of the $k$-integral, $k_{\rm max}$, is the maximum Fourier mode, chosen to avoid unnecessarily heavy computation due to the highly oscillatory nature of the $j_l(k\chi)$. At small scales, we make the Limber approximation \cite{Limber:1954zz} where $j_l(k\chi)\propto \delta(l+1/2-k\chi)$ (where $\delta$ here indicates the Dirac delta), and the power spectrum projection simplifies {to a single integral over $\chi$}. This breaks down on large scales, beyond which we evaluate the full non-Limber integral (we use $l=30$ as the boundary between our Limber and non-Limber calculations).

For $C_l^{v_iv_j}$ and {$C_l^{Xv_j}$, where $X\in\{g_i,\kappa\}$}, we need to project kSZ velocity into harmonic space:
\begin{align}
    C_l^{v_iv_j} &=  \frac{2}{\pi}\int_{0}^{\chi_{\rm dec}} d\chi \int_{0}^{k_{\rm max}}dkk^2\left(\frac{faH}{k}\right)^2 W_{\rm kSZ}^{i}(\chi)j'_l(k\chi) \nonumber \\
    &\int_{0}^{\chi_{\rm dec}} d\chi^\prime W_{\rm kSZ}^{j}(\chi^\prime)j'_l(k\chi')P(k,z,z'),\\
    C_l^{{X}v_j} &= \frac{2}{\pi}\int_{0}^{\chi_{\rm dec}}d\chi\int_{0}^{k_{\rm max}} dk k^2\frac{faH}{k}W_X(k,\chi)j_l(k\chi) \nonumber \\
     & \int_{0}^{\chi_{\rm dec}}d\chi^{\prime} 
     W_{\rm kSZ}^{j}(\chi^\prime)j'_l(k\chi')P(k,z,z'),
\end{align}
where $j_l^\prime(x)$ indicates the derivative of $j_l(x)$ with respect to $x$, which we note can be related to the Bessel functions themselves with the relation
\begin{equation}
    j'_l(k\chi)=\frac{1}{2l+1}[lj_{l-1}(k\chi)-(l+1)j_{l+1}(k\chi)].
\end{equation}

We use these theory spectra in Fisher forecasts for the error bar on $f_{\mathrm{NL}}$; we describe our formalism for the Fisher forecasts in App.~\ref{fisher}. The relevant power spectra in the Fisher matrix include all sources of variance due to noise $N_l$: $C_l^{gg}+N_l^{gg}, C_l^{gv}, C_l^{vv}+ N_l^{vv}, C_l^{g\kappa}$, and $C_l^{\kappa\kappa}+N_l^{\kappa\kappa}$ (where we assume $N_l^{XY}$ is zero for $X\ne Y$), i.e. they include galaxy shot noise $N_l^{gg}$, kSZ velocity reconstruction noise $N_l^{vv}$, and CMB lensing reconstruction noise $N_l^{\kappa\kappa}$. We assume the galaxy shot noise {is given by $N_l^{gg}=1/\bar n$, where~$\bar n$ is the mean galaxy number density}. The kSZ reconstruction noise and the CMB lensing reconstruction noise are sourced (to first order) by the expectation value of the relevant estimators applied to Gaussian fields with no signal (``the $N^{(0)}$ biases''), and so depend on the underlying power spectra of the maps used for reconstruction and how they are weighted; for kSZ these are the temperature power spectrum including all sources of noise $C_l^{TT}+N_l^{TT}$, the galaxy power spectrum $C_l^{gg}+N_l^{gg}$ and their relative weighting via the assumed galaxy-optical-depth correlation $C_l^{g\tau}$, while for CMB lensing it is $C_l^{TT}+N_l^{TT}$ and the CMB polarization power spectra including all sources of noise, i.e., $C_l^{TE}$, $C_l^{EE}+N_l^{EE}$ and  $C_l^{BB}+N_l^{BB}$ for $E$ and $B$ modes respectively. The kSZ velocity reconstruction noise will be discussed in greater detail in Sec.~\ref{ksz recon}.
 
We assume fiducial non-Gaussianity $f_{\rm NL}^{\rm fid}=0$, {fiducial galaxy bias $b_{g,i}^{\rm NL,fid}(z)=0.95/\mathcal{D}(z)$} for LSST \cite{LSSTDarkEnergyScience:2018jkl,LSSTScience:2009jmu} and galaxy bias as specified in Ref.~\cite{Heinrich:2023qaa} for SPHEREx, and {fiducial} velocity biases $b_{v,i}^{\rm fid}$ (we choose only to explore two cases where $b_{v,i}^{\rm fid}=[.33,1]$ with respect to a commonly-used Battaglia-like halo model prediction for the galaxy-electron cross-correlation $C_l^{g\tau}$ \cite{Battaglia:2016xbi} to encompass the range of uncertainty on this parameter). We compute the theoretical spectra in this work, along with the kSZ reconstruction noise, with an adapted version of the \href{https://github.com/jcayuso/ReCCO}{\faGithub\ ReCCO} public code \cite{Cayuso:2021ljq}. 

With these power spectra, we make a Fisher forecast for $\sigma(f_{\rm NL})$, marginalizing over one galaxy and one velocity bias parameter for each redshift bin and assuming Gaussian likelihoods for the spectra. We note that previous forecasts \cite{Munchmeyer:2018eey, Schmittfull:2017ffw} along this line typically assumed $C_l^{gg}$, $C_l^{gv}$ and $C_l^{vv}$ information will be reliable on {the same} scales. Our forecast explores how tightly we can constrain local PNG, if sensitivity to $f_{\rm NL}$ comes solely from $gv$ cross-correlations down to the largest angular scale (although we still include smaller-scale $gg$ to inform $b_g$ and break the $b_g-b_v$ degeneracy in $C_l^{gv}$, and the large-scale $vv$ and $\kappa\kappa$ auto power spectra to aid in CVC). In this cross-correlations approach, $f_{\rm NL}$ is directly constrained by only $gv$ cross correlations at large scales. This is desirable because $gv$ does not suffer from data systematics like $gg$: all additive, uncorrelated noise (or systematic) bias is eliminated in the $gv$ cross correlation, although we note that multiplicative systematics are not eliminated.
We also note that uncorrelated systematics can still add variance to $\sigma(f_{\rm NL})$, which we do not account for in our forecast. We define the angular scale $l_{\rm cut}$ below which we remove all $gg$ auto-information, the \textit{kSZ baseline case} $gg(l>l_{\rm cut})$+$gv$+$vv$ where $b_v^{\rm fid}=0.33$ with respect to our Battaglia-like halo model and SO-like reconstructed velocity, and likewise the \textit{CMB lensing baseline case} $gg(l>l_{\rm cut})$+$g\kappa$+$\kappa\kappa$. The bandpower Fisher formalism, which we review in App.~\ref{fisher}, is particularly suited for this purpose. In addition to the baseline cases, we compare the extent to which $\sigma(f_{\rm NL})$ degrades when removing $vv$ and $\kappa\kappa$, i.e. $gg(l>l_{\rm cut})$+$gv$ and $gg(l>l_{\rm cut})$+$g\kappa$, where $vv$ and $\kappa\kappa$ by themselves do not possess any constraining power on $f_{\rm NL}$ but contribute to CVC. We also define the \textit{all case} when cross correlating all three tracers $gg(l>l_{\rm cut})+gv+vv+g\kappa+\kappa\kappa+\kappa v$ and correspondingly $gg(l>l_{\rm cut})+gv+g\kappa+\kappa v$.

\subsection{kSZ Velocity Reconstruction}\label{ksz recon}

We work with a curved-sky formalism for kSZ velocity reconstruction, using a convolution of small-scale $\delta_g$ and CMB temperature. {The kSZ temperature $T_{\text{kSZ}}(\hat n)$ is given by
\begin{align}
    \frac{T_{\text{kSZ}}(\hat n)}{\bar T} =-\sigma_T \int a(\chi)n_e(\chi,\hat{n}) {v_r}(\chi,\hat{n}) d\chi,
\end{align}
where $\bar T= 2.725$K is the CMB mean temperature \cite{Fixsen:1996nj}, $v_r$ {is the radial peculiar velocity}, and $\sigma_T$ is the Thomson scattering cross section.} The differential optical depth is defined as $\dot{\tau}(\chi,\hat n)=\sigma_T a(\chi)n_e(\chi,\hat n)$, and its uncertainty leads to a degeneracy with $v_r$, the consequences of which are explained further in Sec.~\ref{bv_section}\fionaedit{.} In App.~\ref{ksz_estim}, we review the optimal kSZ velocity quadratic estimator \cite{McCarthy:2024nik, Deutsch:2017ybc, Cayuso:2021ljq, Bloch:2024kpn} given by
\begin{equation}\label{vel_estim}
    \hat{v}_{LM}^i = A_L^i\sum_{l m; l' m'}(-1)^M \begin{pmatrix} l & l' & L \\ m & m' & -M \end{pmatrix} \omega^i_{l l' L}T_{l m}\delta ^{g^i}_{l' m'},
\end{equation}
where 
\begin{align}
    \omega^i_{ll'L} &= \frac{f_{ll'L}C_{l'}^{\tau g_i}}{\hat{C}_{l}^{TT}\hat{C}_{l'}^{g_ig_i}};\\
    A_L^{i} &= (2L + 1) \left( \sum_{l; l'} \frac{\left( f_{l l' L} C_{l'}^{g_i\tau} \right)^2}{\hat{C}_{l}^{TT}\hat{C}_{l'}^{g_ig_i}} \right)^{-1}\\
    f_{ll'L}&=\sqrt{\frac{(2l + 1)(2l' + 1)(2L + 1)}{4\pi}} \begin{pmatrix} l & l' & L \\ 0 & 0 & 0 \end{pmatrix}.
\end{align} 
Here, $\hat{C}_l^{g_ig_i}$ and $\hat{C}_l^{TT}$ are the galaxy and CMB temperature autospectra inclusive of all sources of noise and foregrounds, and $C_l^{g_i\tau}$ is the  galaxy-optical-depth correlation. In a real analysis, $\hat C_l^{g_ig_i}$ and $\hat C_l^{TT}$ could come from measuring autospectra of the $T^{\rm kSZ}$ and $\delta_g$ fields used for kSZ velocity reconstruction. For forecasting the reconstruction noise, we use their theoretical predictions along with the predicted shot noise and foreground and instrumental noise for the galaxy and CMB surveys. $C_l^{g_i\tau}$ comes from a model, which can introduce a bias if incorrect (this bias is incorporated by the $b_v$ parameter).

Taking the autospectrum of the kSZ velocity estimator given in Eq.~\eqref{vel_estim} yields a noise bias component, i.e. an expectation value of the estimator variance in the absence of kSZ signal, which must be appropriately accounted for in the Fisher matrix we use for forecasting. Heuristically, the zeroth-order noise comes from the $\langle TT\rangle\langle\delta_g^i\delta_g^j\rangle$ contribution to the velocity autospectrum $\langle v_iv_j\rangle$. This is due to our assumption that the non-kSZ contribution to $T$ and~$\delta_g$ are uncorrelated, Gaussian random fields at leading order. We use Wick's Theorem to break down $\langle T\delta_g^iT\delta_g^j\rangle$, where cross-terms (i.e., those containing $\langle T\delta_g^i\rangle$) vanish.\footnote{{If we drop the assumption that the non-kSZ contributions to $T$ and $\delta_g$ are uncorrelated, terms containing the Gaussian $\left<T\delta\right>$ correlations (e.g., those induced by correlations between the galaxy distribution and foregrounds to the CMB) would also be included; the full estimator including these terms was presented in Ref.~\cite{Cayuso:2021ljq}. Corrections to the estimator are of order of the correlation coefficient between the galaxies and the temperature; in previous data analyses these small but non-zero correlations have been neglected.}} Manipulating Wigner algebra, we find the zeroth-order reconstruction noise:
\begin{equation}\label{reconnoise}
    N_L^{v_iv_j} = \frac{A_L^iA_L^j}{2L+1}\sum_{ll'}\omega^i_{ll'L}\omega^j_{ll'L}\hat{C}_{l}^{TT}\hat{C}_{l'}^{g_ig_j}
\end{equation}
which reduces to $N_L^{v_i v_i}=A_L^i$ when correlating $v$ within the same bin. Since $v$ reconstruction is dominated by small scales $l, l'\gg L$ where $\omega^i_{ll' L}\omega^j_{ll' L}\propto f^2_{ll' L}\propto (2L+1)$, $N_L^{ij}$ is approximately white over the range $L\leq l_{\rm max}=200$ used in our forecasts.

\subsection{Impact of the kSZ Optical Depth Degeneracy on $f_{\rm NL}$ Constraints}\label{bv_section}

KSZ velocity reconstruction leads to a scale-independent bias known as the ``kSZ velocity bias'' given by 
\begin{equation}
    b_v^i\simeq\sum_l\frac{(2l+1)C_l^{g_i\tau}C_{l,\rm true}^{g_i\tau}}{\hat{C}_l^{TT}\hat{C}_l^{g_ig_i}}/\sum_l\frac{(2l+1)(C_l^{g_i\tau})^2}{\hat{C}_l^{TT}\hat{C}_l^{g_ig_i}}
\end{equation}
where $C_l^{g_i\tau}$ is the galaxy-electron power spectrum assumed in the reconstruction and $C_{l, \rm true}^{g_i\tau}$ describes the true underlying galaxy-electron correlation.

Various separate measurements of the velocity bias, such as with DESI luminous red galaxies (LRGs)~\cite{McCarthy:2025brx, Hotinli:2025tul, Lai:2025qdw, Hadzhiyska:2025egz}, have shown increasing evidence for $b_v<1$ compared to the kSZ velocity signal amplitude expected from AGN profiles like Battaglia~\cite{Battaglia:2016xbi}, where the velocity bias is largely measured to be between $0.2$ and $0.5$ at redshifts $z<1$ \cite{Hadzhiyska:2025egz,Chaussidon:2026vmn}. For our purposes, we assume $b_v^{\rm fid}=0.33$ as a middle ground. {As pointed out in Ref.~\cite{Tishue:2025cvp}, this low value of $b_v$ reduces $C_l^{gv}$ and $C_l^{vv}$ by factors of $\sim 3$ and $\sim 9$ respectively, and hence the $f_{\rm NL}$ constraining power}. We compare its impact on $gv$  cross correlations with $b_v^{\rm fid}=1$~\cite{Tishue:2025cvp}. The kSZ velocity SNR determines the impact of low $b_v^{\rm fid}$ on the $f_{\rm NL}$ constraint, which we argue is suppressed for $C_l^{vv}/N_l^{vv}\gg 1$ when doing cross correlations alone. A heuristic argument for this is given in App.~\ref{snr_estimate}.

\begin{figure}
    \raggedright
    \includegraphics[width=1.1\linewidth]{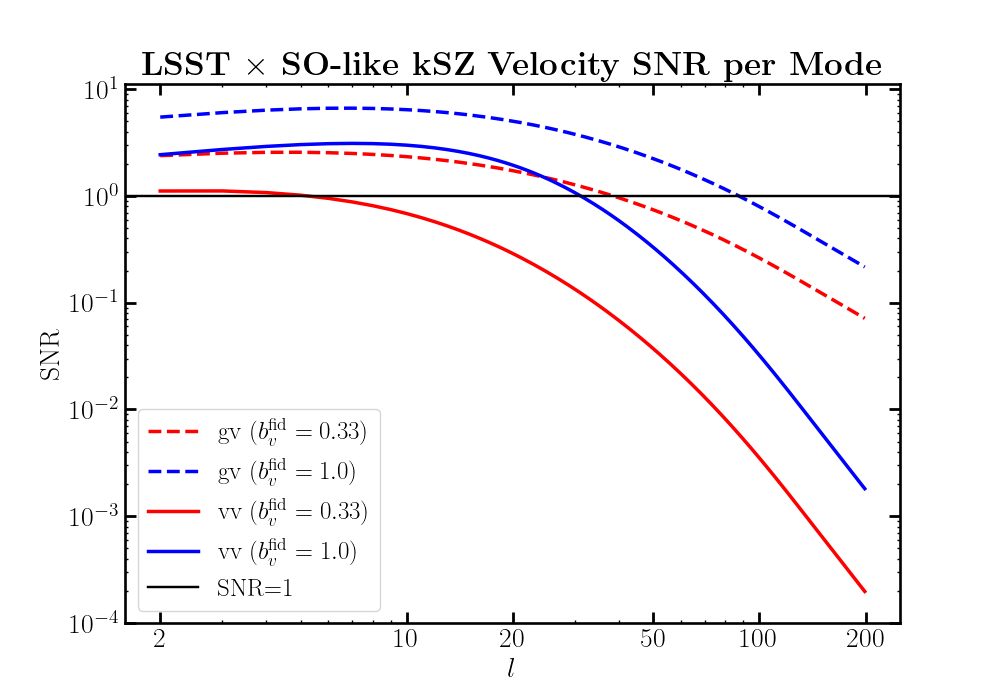}
    \caption{The forecast signal-to-noise ratio (SNR) of $C_l^{gv}$ and $C_l^{vv}$ per angular mode $l$, for an SO+LSST-like setup, for different fiducial assumptions of $b^{\mathrm{fid}}_v$. A smaller kSZ signal (indicated by $b_v^{\rm fid}=0.33$) reduces all $v$-related signals.}
    \label{fig:SNR}
\end{figure}

In general, the SNR of $gv$ is larger than that of $vv$. In Fig.~\ref{fig:SNR}, we show our forecast for the SNR per mode of $gv$ and $vv$ for the SO+LSST-like set up. We see that the $vv$ measurement for $b_v^{\rm fid}=0.33$ will barely be signal-dominated even at large scales {when we consider velocity reconstructed assuming SO-like noise levels and $\delta_g$ appropriate for LSST}, {thus} contributing {essentially} no CVC. We note that $b_v^{\rm fid}=0.33$ was measured for low-redshift DESI LRGs \cite{McCarthy:2025brx}, ELGs, and QSOs \cite{Chaussidon:2026vmn} but its value at higher redshift bins remains to be determined.

\section{Cross Correlation Forecasts }\label{forecasts}

In our cross correlation forecast, we assume post-foreground-cleaning SO-like noise levels~\cite{SimonsObservatory:2018koc} for the kSZ velocity reconstruction, obtained from public SO-like noise models.\footnote{\url{https://github.com/simonsobs/so_noise_models/tree/master/LAT_comp_sep_noise/v3.1.0}} For galaxies, we assume either LSST-like or SPHEREx-like galaxy number density and photo-redshift errors, and separately consider the cross-correlation of $\delta_g$ in each case with velocity. We calculate the kSZ velocity reconstruction noise with Eq.~\eqref{reconnoise}, using theoretical predictions for small-scale LSST $\hat C_l^{gg}$ (signal computed with the halo model + shotnoise), and a prediction for the CMB temperature power spectrum $\hat C_l^{TT}$ obtained from a prediction for a standard Internal Linear Combination (ILC), minimum-variance (MV) cleaning estimator:  it consists of signal, residual foregrounds, and SO-like instrumental noise, assuming ``goal" sensitivity of the survey and sky coverage $f_{\rm sky}=40\%$. Fig.~\ref{fig:act_v_so_cltt} shows the lower noise levels SO is forecasted to reach compared to ACT, in the CMB temperature. We explain further survey-specific details in Secs.~\ref{lsstxso}~-~\ref{spherexso}.

\begin{figure}
    \centering
    \includegraphics[width=\linewidth]{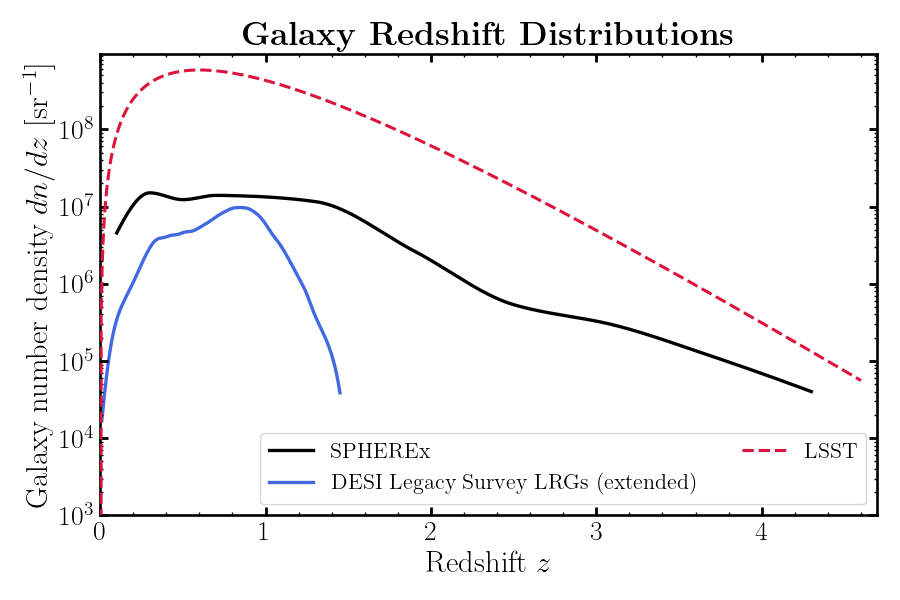}
    \caption{Absolute galaxy number density distribution in redshift space for forecasted LSST, forecasted SPHEREx, and the DESI Legacy extended LRGs sample \cite{Zhou:2023gji}. The SPHEREx and DESI curves are interpolated from binned values and smoothed via linear interpolation.}
    \label{fig:dndz}
\end{figure}

\subsection{Galaxy $\times$ kSZ Velocity}
\subsubsection{LSST $\times$ SO-like}\label{lsstxso}

For LSST, we use the specific $dn/dz$ outlined in Ref.~\cite{LSSTScience:2009jmu} and shown in Fig.~\ref{fig:dndz}:
\begin{align}\label{lsst_dndz}
    \frac{dn}{dz} &= \frac{N}{2z_0}\left(\frac{z}{z_0}\right)^2 \exp\left(-\frac{z}{z_0}\right),
\end{align} 
where we assume the total galaxy angular number density $N=55\,\mathrm{arcmin^{-2}}$ (appropriate for the LSST Gold galaxy sample) and $z_0=0.3$. We assume that the photometric redshift error scales with redshift as $\frac{\sigma(z)}{1+z}=0.05$. We divide redshift space in the region $0.02\leq z\leq 7$
into 18 bins (a number which we have found saturates the $f_{\rm {NL}}$ constraint, see App.~\ref{binning}), with boundaries in photo-$z$ given by $[0.02, 0.135, 0.25, 0.375, 0.5, 0.625, 0.75, 0.875, 1, 1.25, 1.5$, $1.75, 2, 2.25, 2.5, 3, 3.5, 4, 7]$. The redshift-space binning is uneven due to fewer galaxies being detectable at higher redshifts, and we marginalize analytically over $b_g^i$ \footnote{{While the calculation of the $C_l^{g_iX}$ signals assume continuous fiducial galaxy bias integrated over redshift, for the analytic Fisher derivatives we model this as $C_l^{g^iX}=b_g^iC_l^{g^iX}/(0.95/\mathcal{D}(z_{\mathrm{mid}}^i))$, $b_g^i=0.95/\mathcal{D}(z_{\mathrm{mid}}^i)$ where $z_{\mathrm{mid}}^i$ indicates the midpoint of photo-$z$ bin $i$, and we analytically differentiate by  $b_g^i$}}. We assume that shot noise $N^{gg}_{ij}\propto\delta_{ij}$ is diagonal. To compute $N^{gg}_{ij}$, we integrate over the redshift range of $dn_i/dz$ in Eq.~\eqref{eq_dnidz}.

\begin{figure*}
    \centering
    \includegraphics[width=0.49\textwidth]{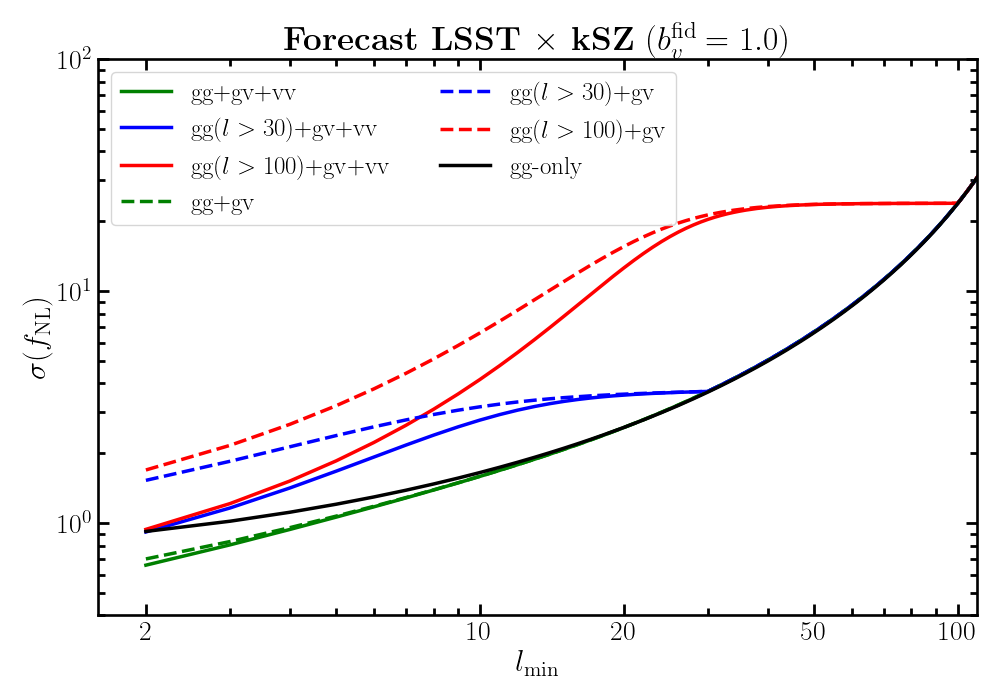}    \includegraphics[width=0.49\textwidth]{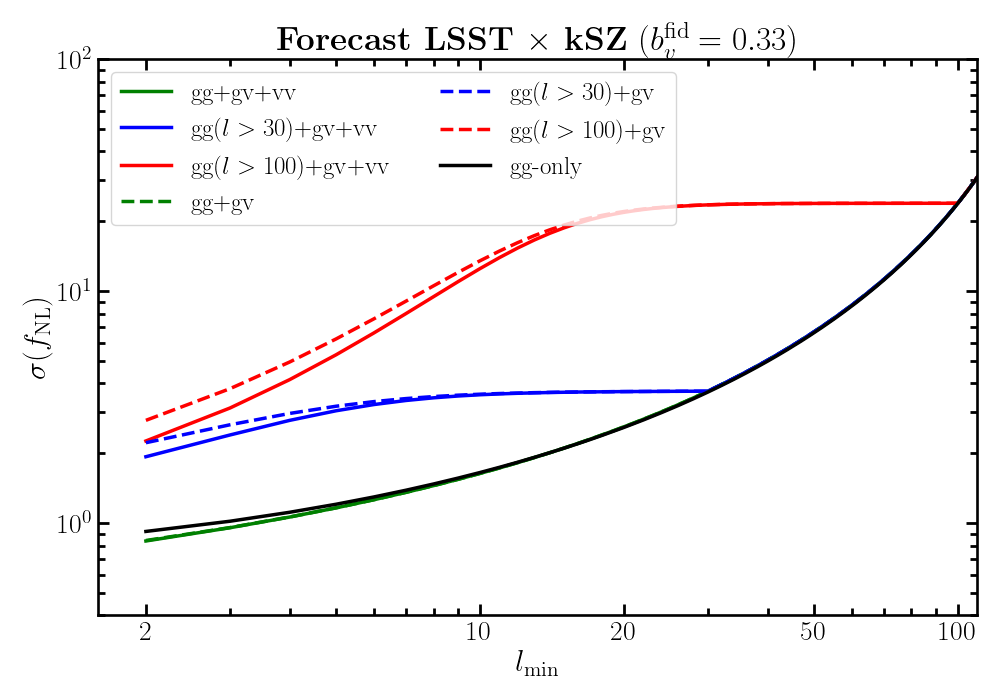}
    \caption{{\textit{Left}: LSST$\times$SO-like $\delta_g\times v$ forecast assuming $b_v^{\rm fid}=1$. Full-scale $gg+gv+vv$ yields $\sigma(f_{\rm NL})\sim 0.65$ and $gg(l>30, 100)+gv+vv$ yield $\sigma(f_{\rm NL})=0.92, 0.95$ at $l_{\rm min}=2$. Compared to $gg$-only, adding $gv+vv$ reduces the forecasted error by $26\%$. With $gg+gv$, constraints deteriorate only slightly but worsen to $\sim 1.5$ and $\sim 1.7$ for $gg(l>30)+gv$ and $gg(l>100)+gv$. \textit{Right}: LSST$\times$SO-like $\delta_g\times v$ baseline forecast assuming $b_v^{\rm fid}=0.33$. Comparing with matching scenarios to the left, constraints deteriorate due to a weaker $v$ signal. For $gg+gv+vv$ and $gg+gv$, $\sigma(f_{\rm NL})$ increases to $\sim 0.85$ at $l_{\rm min}=2$, compared to $b_v^{\rm fid}=1$. For $gg(l>30, 100)+gv+vv$, $\sigma(f_{\rm NL})$ rises from $\sim 0.95$ to approximately $1.95$ and $2.3$ respectively. Overall, even in the case of no-$vv$ and low $b_v^{\rm fid}$, relying on $gv$ cross correlations forecasts $\sigma(f_{\rm NL})\sim 2.2$.}}
    \label{lsst_bv_33}
\end{figure*}

{Converting number density from units of arcmin$^{-2}$ into steradian$^{-1}$, the shot noise in bin $i$ is given by }$N_{gg}^i=1/\int_{z=0}^{\infty} (dn_i/dz)dz$. Forecast $1\sigma$ $f_{\rm NL}$ constraints are shown in Fig.~\ref{lsst_bv_33} as a function of the minimum multipole included $l_{\mathrm{min}}$, for $b_v^{\mathrm{fid}}=1$ and $b_v^{\mathrm{fid}}=0.33$. With $b_v^{\rm fid}=1$ and baseline $l_{\rm cut}=30$, we forecast $\sigma(f_{\rm NL})=0.92$ if we can reach $l_{\mathrm{min}}=2$. As expected, the baseline $b_v^{\mathrm{fid}}=0.33$ degrades $\sigma(f_{\mathrm{NL}})$ noticeably with respect to the case where $b_v^{\mathrm{fid}}=1$ due to lower SNR (Fig.~\ref{fig:SNR}), with marginal CVC contributions from $vv$. For reference, our $gg$-only forecast is $\sigma(f_{\rm NL})\simeq 0.9$ at $l_{\rm min}=2$ for $f_{\rm sky}=0.4$, whereas Ref.~\cite{Munchmeyer:2018eey} finds $1.35$,  assuming the universal halo mass function but lower $\delta_c=1.42$ and $f_{\rm sky}=0.3$.  Ref.~\cite{Tishue:2025cvp} finds $1.37$ {also with $f_{\rm sky}=0.3$}. Scaled as $f_{\rm sky}^{-1/2}$ to $f_{\rm sky}=0.4$, both correspond to $\simeq 1.2$. We note though that we assume a higher number density, $55$ against $40\,{\rm arcmin}^{-2}$.

In Fig.~\ref{lsst_bv_33}, we observe $\sigma(f_{\rm NL})$ flattening out after removing large-scale $gg$, until reaching much larger angular scales when the SNR of $gv$ reaches $\gtrsim\mathcal{O}(1)$ (in accordance with the SNR in Fig.~\ref{fig:SNR}) and begins contributing to the constraint. The dependence of $\sigma(f_{\rm NL})$ on $l_{\rm min}$ shows that $gg$ provides the primary constraining power at these smaller scales, and variance cancellation from the large-scale velocity autospectrum ($l_{\rm  min}\lesssim 10$) is appreciable if signal dominated (e.g. $b_v^{\rm fid}=1$). Note that we have accounted for partial sky coverage only by an $f_{\mathrm{sky}}$ factor (where $f_{\mathrm{sky}}$ is the fraction of sky we expect to observe), and have not accounted for any difficulties in measuring low-$l$ modes due to survey geometries.

\subsubsection{SPHEREx $\times$ SO-like and ACT-like}\label{spherexso}

We also explore forecasts for a SPHEREx+SO-like setup. SPHEREx is quasi-spectroscopic, measuring a low-resolution spectrum in 102 channels over $0.75–5 \mu$m for every source on the sky. It gives a flux-limited, complete sample like a photometric survey, but with redshift errors as small as $0.003$, which preserve radial modes up to $0.1 h$ Mpc$^{-1}$. {This gives the SPHEREx samples much better redshift precision than LSST, which observes in six broad bands.}

Following Ref.~\cite{Heinrich:2023qaa}, we assume that the SPHEREx galaxies can be split into 5 different samples which we label with indices $A=1,2,3,4,5$, with different photo-redshift errors {given by $\sigma(z)/(1+z)=0.003, 0.01, 0.03, 0.1, 0.2$} respectively, and each with different number densities and linear biases. We also assume that kSZ velocity will be reconstructed individually for each sample. 

The tomographic SPHEREx $dn_i/dz$ uses the $\sigma(z)$-convolution in Eq.~\eqref{eq_dnidz}, where we first interpolate over the per-bin number density to get the smoothly distributed SPHEREx $dn/dz$ in Fig.~\ref{fig:dndz}. When calculating the velocity reconstruction noise, which requires small-scale $C_l^{g_i\tau}$, we assume for simplicity the same halo occupation distribution (HOD) parameters as LSST. There are eleven $z$-bins following \cite{Heinrich:2023qaa}\footnote{{We use the pre-launch forecast of number densities and biases from the public SPHEREx products repository, as tabulated in \cite{Heinrich:2023qaa}. The cosmology sample selection is still being finalized with flight data \cite{Bock:2025ijl, Huai:2025prs}.}} containing 3D galaxy modes in SPHEREx, with boundaries at redshifts $[0,0.2,0.4,0.6,0.8,1,1.6,2.2,2.8,3.4,4,4.6]$, but we explore finer binning to saturate the constraining power on $f_{\rm NL}$. 

\begin{figure}
    \centering
    \includegraphics[width=\linewidth]{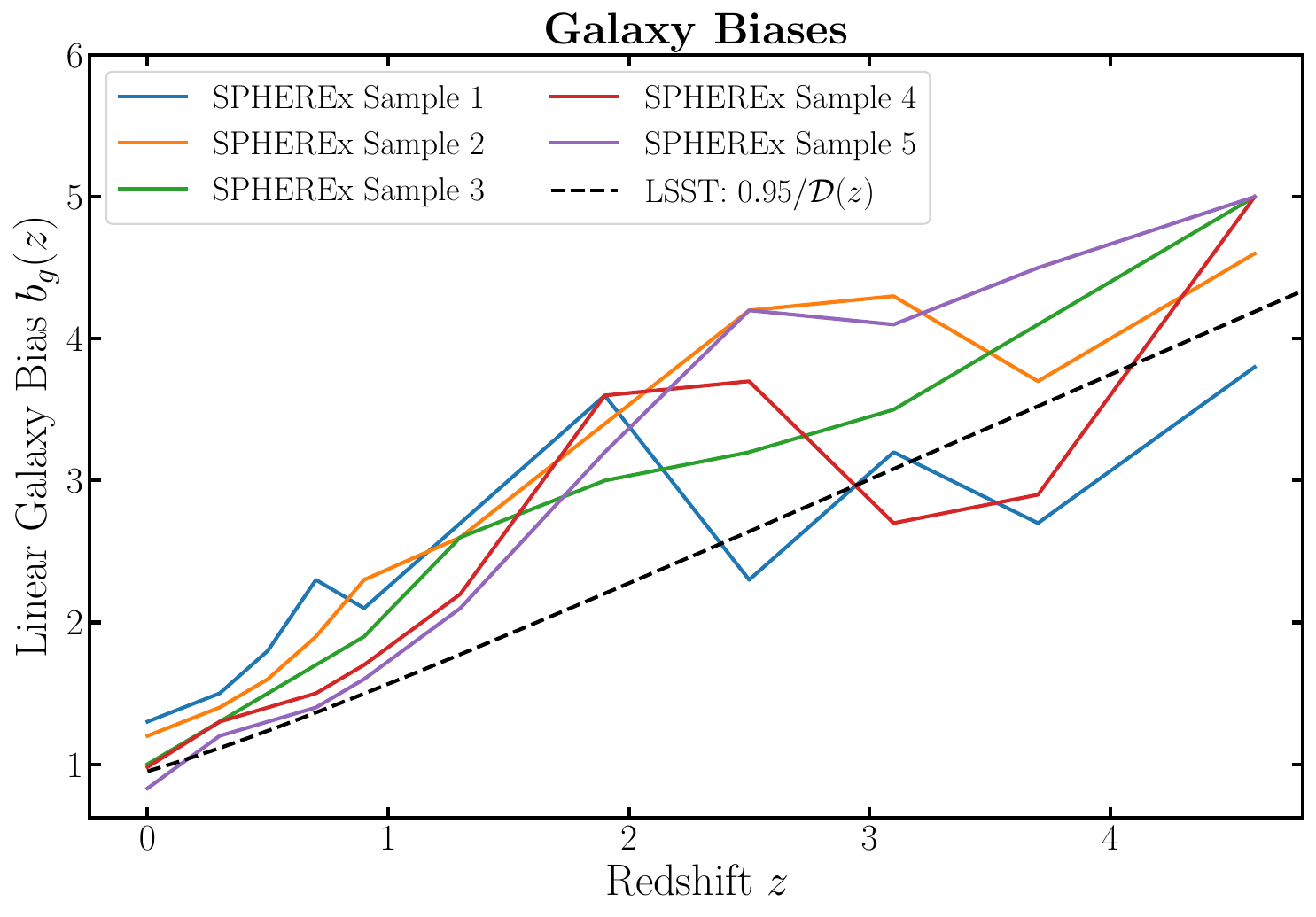}
    \caption{Fiducial galaxy biases linearly interpolated over redshift for all five galaxy samples for SPHEREx \cite{Heinrich:2023qaa}, along with fiducial galaxy biases for LSST.}
    \label{fig:interpolated_redshifts}
\end{figure}

We further note the different SPHEREx galaxy bias. Whereas we model LSST with $b_g^{\rm fid}(z)=0.95/\mathcal{D}(z)$, for SPHEREx we use the galaxy bias predictions shown {in Fig.~\ref{fig:interpolated_redshifts} plotted from exact values given by~\cite{Heinrich:2023qaa}}. Furthermore, each SPHEREx galaxy sample is a distinct but correlated tracer of the LSS and cross-correlating them with each other will lead to CVC. This is in addition to potential CVC from cross-correlating $\delta_g$ with the kSZ velocity field, so we fully exploit this internal multitracer technique. 

For 2D redshift tomography, when working with many redshift bins and approaching the {bin width $\Delta z \rightarrow \sigma(z)$}  limit, the constraining power on $f_{\rm NL}$ is completely saturated. For quasi-spectroscopic surveys like SPHEREx (and photometric surveys like LSST), the number of tomographic bins $N$ required for saturation depends on the $\sigma(z)$ of individual samples: binning redshift space such that {bin width} $\lesssim\sigma(z)$  will cease to improve the constraint, as such surveys are fundamentally limited by this redshift measurement uncertainty. In practice, this limit is saturated before we reach $\sigma(z)$ in bin width. For SPHEREx, we empirically find the constraint-saturating $N^A_{\rm bins}=44,44,33,22,22$ for samples $A=1,2,3,4,5$ such that binning redshift space more finely ceases to decrease forecasted $\sigma(f_{\rm NL})$. For purposes of this forecast, the default eleven $z$-bins and the per-bin number densities are evenly divided by $N_{\rm bins}^X/11$. We interpolate over the 11 galaxy biases {for each of the five samples against redshift} to obtain fiducial biases for each subdivided bin{, as shown} in Fig.~\ref{fig:interpolated_redshifts}.

Thus, for SPHEREx we work with $C_l^{X^A_iY^B_j}$, where $A$ and $B$ label the sample indices, $i,j$ the redshift bins, and $X, Y$ the tracers ($\delta_g$, $v$, or {$\kappa$}). {In this context, ``cross correlations alone'' means we discard $C_l^{g_i^Ag_j^B}(l<l_{\rm cut})$ for all $A,B$ (including all intersample correlations) and $i,j$.} We assume the distribution of each galaxy sample is Poissonian and uncorrelated between the samples (which may be a simplification), so shot noise $N_l^{g^A_ig^B_j}\propto \delta_{ij}\delta_{AB}$, where $\delta_{ab}$ indicates the Kronecker $\delta$. The intersample reconstruction noise is in general nonzero due to the correlations between galaxies in different samples but at the same redshift (due to the overlapping window functions induced by the photo-$z$ error). However, since kSZ velocity is reconstructed from small-scale  $(l\gtrsim 2000)$ $\delta_g$ and $T$, where shot noise dominates over the galaxy autospectrum by orders of magnitude, the impact of intersample reconstruction noise is subdominant {to that from the same sample}, which has a large contribution from shot noise. We note this SPHEREx forecast assumes $f_{\rm sky}=0.75$ for $gg$ but $f_{\rm sky}=0.4$ for $gv$ and $vv$. Specifically, we assume the observed sky consists of two independent patches: the SO-SPHEREx overlapping patch with $f_{\rm sky}=0.4$ and the SPHEREx-only patch with $f_{\rm sky}=0.35$. Thus, the total Fisher matrix $\mathcal{F}$ is a combination of Fisher matrix in the overlapping patch $\mathcal{F}^{gg+gv+vv}$ and the additional patch $\mathcal{F}^{gg}$, $\mathcal{F}=0.4\mathcal{F}^{gg+gv+vv}+0.35\mathcal{F}^{gg}$. We note there could be additional correlations between galaxy modes outside (inside) of the SO-SPHEREx overlapping sky patch with velocity modes within (outside), which we have assumed uncorrelated in our forecast. These correlations represent additional modes that we can measure and lower noise that could reduce $\sigma(f_{\rm NL})$ \cite{Xu:2023qmp}. 

\begin{figure*}
    \centering
    \includegraphics[width=0.49\textwidth]{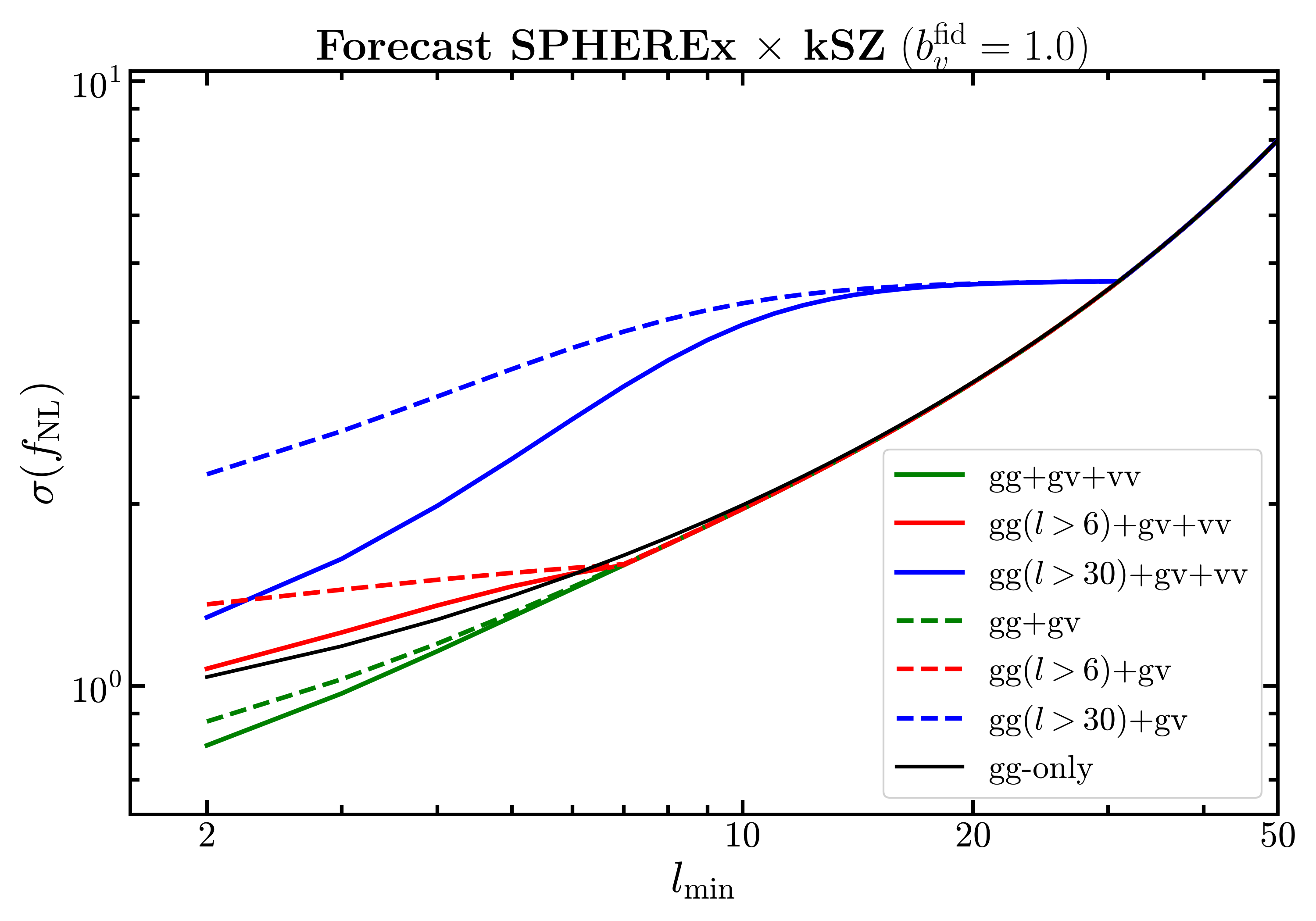}    \includegraphics[width=0.49\textwidth]{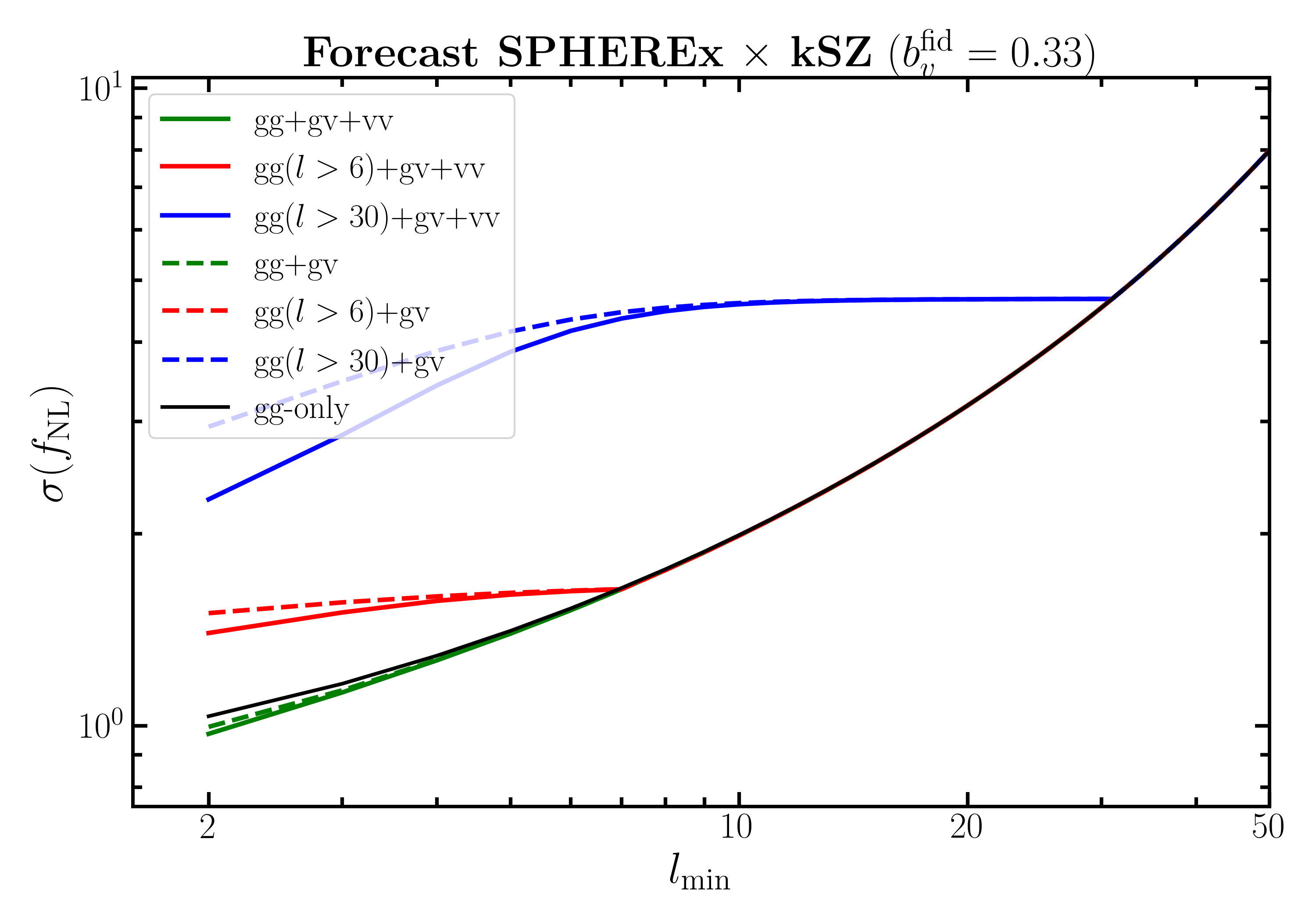}
    \caption{\textit{Left}: SPHEREx $\times$ SO-like kSZ ($\delta_g\times v$) forecast, assuming $b_v^{\rm fid}=1.0$. SPHEREx gg-only gives $\sigma(f_{\rm NL})\sim 1.5$ at {a large angular scale target like $l_{\rm min}=6$}. We highlight that $gg(l>6)+gv+vv$ gives $\sigma(f_{\rm NL})=1.07$, approaching the $\sigma(f_{\rm NL})\sim 1$ threshold, with significant variance cancellation from the $vv$ autospectrum. \textit{Right}: SPHEREx $\times$ SO-like kSZ ($\delta_g\times v$) forecast, assuming $b_v^{\rm fid}=0.33$. Cross correlation alone $gg(l>6)+gv+vv$ only slightly reduces the error bar to $\sigma(f_{\rm NL})\sim 1.4$. Without $vv$ variance cancellation, $gg(l>6)+gv$ barely contributes to the constraint.}
    \label{fig:spherex_multi_bv33}
\end{figure*}

The forecasted errorbar $\sigma(f_{\rm NL})$ from cross-correlating velocity with SPHEREx galaxies is shown in Fig.~\ref{fig:spherex_multi_bv33} for $b_v^{\rm fid}=1$ and $0.33$ {(as reference, our $gg$-only forecast with $l_{\rm min}=2$ gives $\sigma(f_{\rm NL})=1.04$; three-dimensional power-spectrum forecasts for the same sky fraction give $0.87$~\cite{SPHEREx:2014bgr} and $0.78$~\cite{Wen:2026mpw}).} The behaviour of  $\sigma(f_{\rm NL})$ against $l_{\rm min}$ for SPHEREx tells a similar story as for LSST, but we find the constraint driven by not only signal-dominated, large-scale velocity modes ($l_{\rm min}\lesssim 10$) cross correlated with galaxies, but also significant variance cancellation from the velocity autospectrum. 

\begin{figure}[h]
    \centering
    \includegraphics[width=\linewidth]{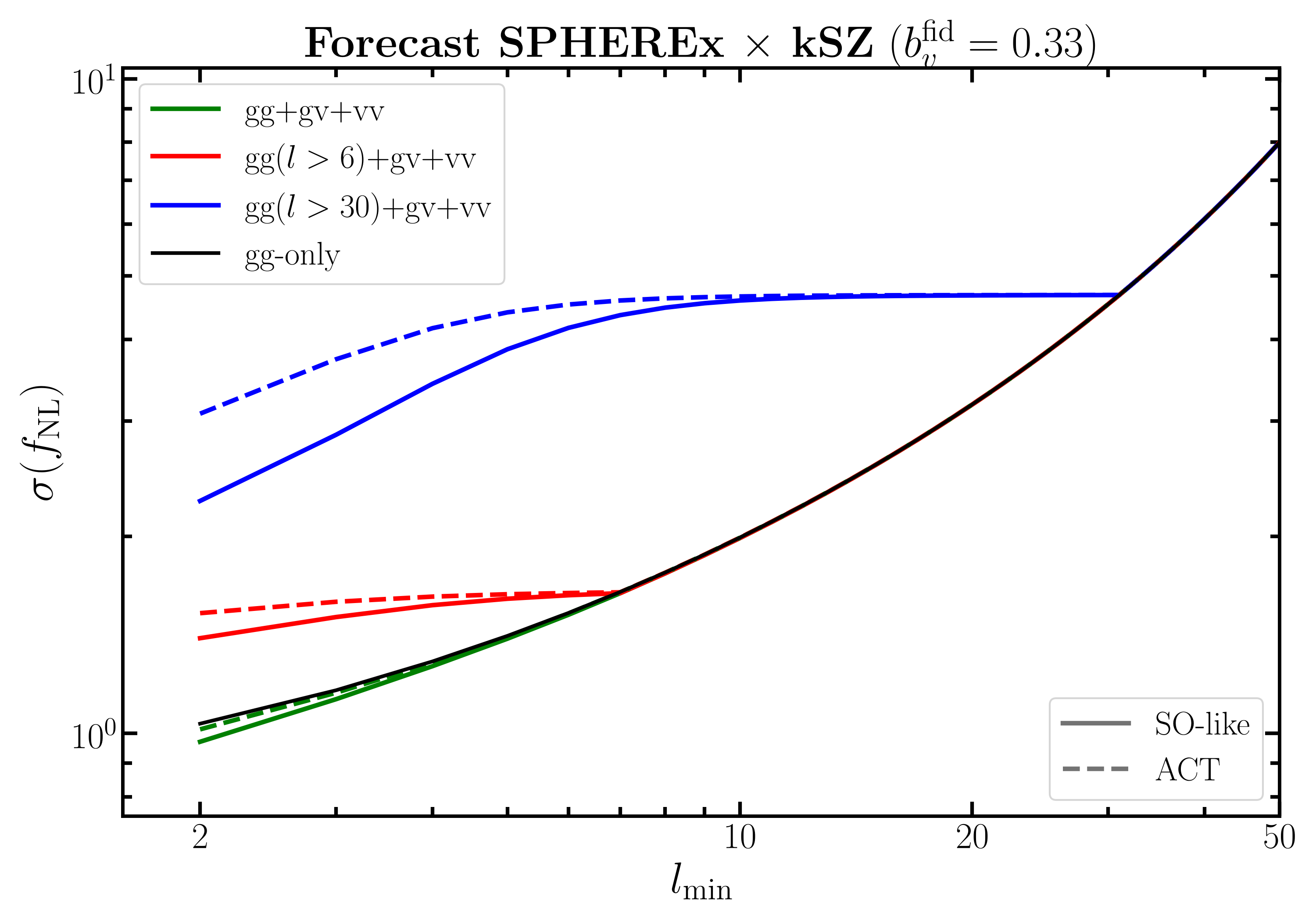}
    \caption{SPHEREx $\times$ kSZ forecast with multitracer galaxy and velocity, comparing between an ACT vs SO-like reconstruction noise: lower noise of the latter due to cleaner $\hat{C}_l^{TT}$ (as shown in Fig.~\ref{fig:act_v_so_cltt}) tightens the $f_{\rm NL}$ constraint. {We assume $f_{\rm sky}=0.75$ for SPHEREx $gg$, while $f_{\rm sky}=0.4$ for $gv$ and $vv$ in both ACT and SO.}}
    \label{spherex_act_so}
\end{figure}

We also forecast the $f_{\mathrm{NL}}$ constraints from cross-correlating SPHEREx with the Atacama Cosmology Telescope (ACT)~\cite{2016JLTP..184..772H,2025JCAP...11..061N,2024PhRvD.109f3530C}. The ACT temperature map has $12\mu$K-arcmin {instrumental} noise level, in comparison to $6\mu$K-arcmin for SO \cite{SimonsObservatory:2018koc} and fewer frequency channels available for foreground variance removal. For purposes of this forecast, this makes a difference in the reconstruction noise, where the {kSZ estimate} for SO is less noisy compared to ACT (Fig.~\ref{fig:act_v_so_cltt}, {as quantified by $\hat C_l^{TT}$}). In Fig.~\ref{spherex_act_so}, we show the expected degradation in constraining power when going from SO to ACT.  

\begin{figure}
    \centering
    \includegraphics[width=\linewidth]{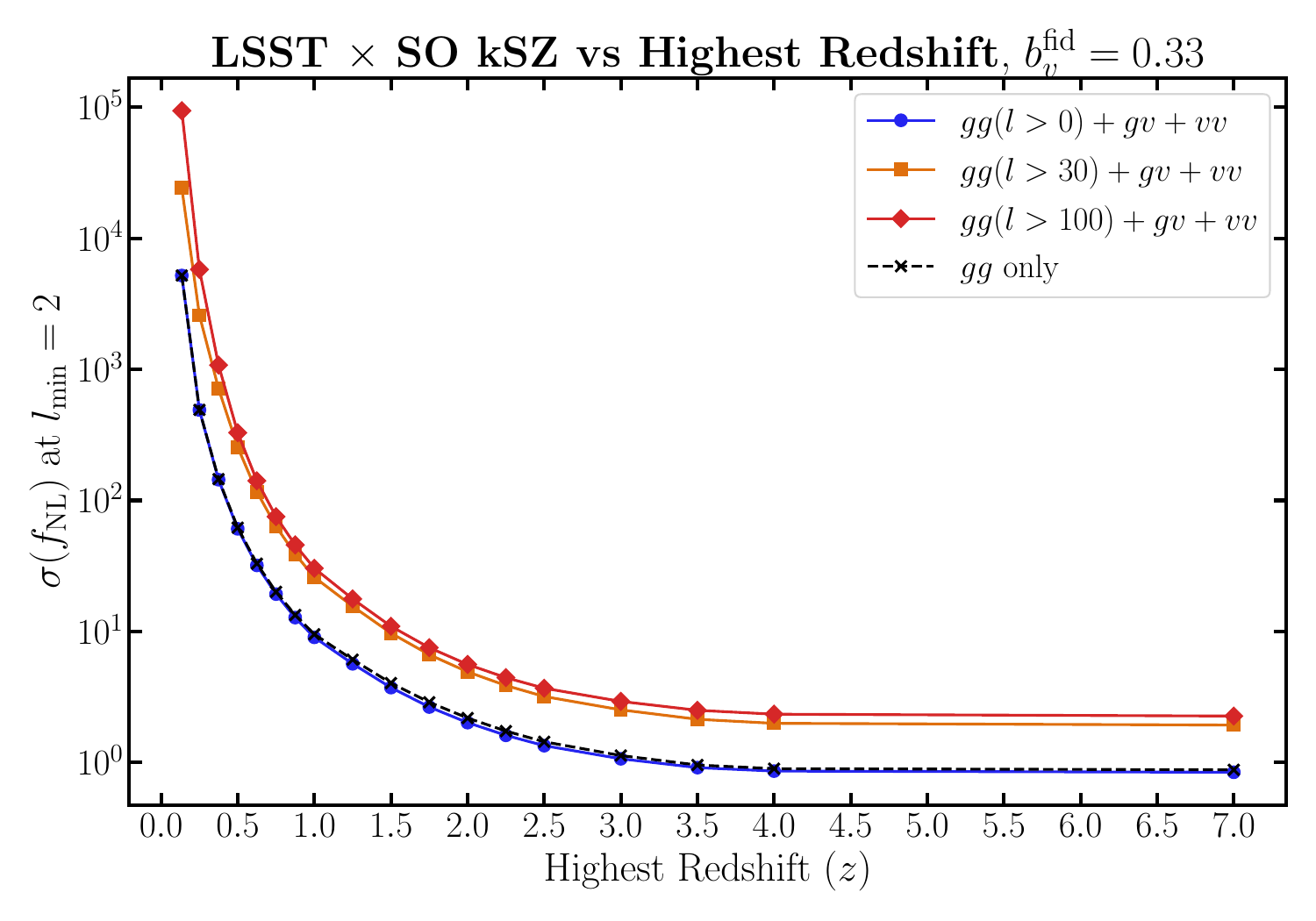}
    \caption{The dependence of $\sigma(f_{\rm NL})$ on the highest redshift information used. In this figure, we drop the highest, second-highest, ... redshift bins, going from right to left on the horizontal redshift axis. Using LSST $\times$ SO-like kSZ as an example, we find that {the} cross correlation alone {forecast is $\sigma(f_{\rm NL})\sim 25$ when the highest redshift is $z=1$. Thus, higher redshifts contribute significantly to the constraint.}}
    \label{redshift_sigma}
\end{figure}

\subsubsection{From DESI LRGs $\times$ACT to LSST/SPHEREx$\times$SO}\label{improvement}

We discuss here the main factors that take cross-correlation-only constraints from their current level to $\sigma(f_{\rm NL})\sim\mathcal{O}(1)$ for next-generation LSS$\times$CMB surveys. At present, the DESI DR1 power spectrum of spectroscopic LRGs and quasars gives $f_{\rm NL}=-3.6^{+9.0}_{-9.1}$ from galaxy clustering alone~\cite{Chaussidon:2024qni}, while analyses that use kSZ velocities reconstructed with ACT, without the galaxy auto-spectrum, reach $\sigma(f_{\rm NL})\simeq 35$~\cite{Hotinli:2025tul,Chaussidon:2026vmn}.

Our forecast for photometric DESI LRGs with ACT (App.~\ref{desi}) agrees with this level. With $f_{\rm sky}=0.195$, cross-correlations alone give $\sigma(f_{\rm NL})\simeq 21$--35, with and without $vv$; scaled as $f_{\rm sky}^{-1/2}$ to the 8--13\% of the sky analysed in Refs.~\cite{Hotinli:2025tul,Chaussidon:2026vmn}, this becomes $\simeq 25$--53. Restricted to $l\geq 6$ and without $vv$, as in Ref.~\cite{McCarthy:2025brx}, it gives 38--84, against the $f_{\rm NL}=-180^{+61}_{-86}$ found there.
Galaxy clustering alone, in contrast, falls well short of its forecast. For the same setup, we forecast $\sigma(f_{\rm NL})\simeq 10$, or $\simeq 7.5$ when scaled to the 14{,}000\,deg$^2$ of photometric DESI LRGs analysed in Ref.~\cite{Rezaie:2023lvi}. That analysis finds $f_{\rm NL}=34^{+24}_{-44}$, an interval three to six times wider, and cleaning with nine imaging maps rather than three more than doubles its uncertainty. Much of the large-scale clustering information is thus lost to imaging systematics, which motivates the cross-correlation-alone approach in this work.
% --- END NEW ---

Three factors close the gap from $\sigma(f_{\rm NL})\simeq 35$ to $\mathcal{O}(1)$. The first is redshift reach: Fig.~\ref{redshift_sigma} shows that adding galaxies at $1<z<4$ reduces $\sigma(f_{\rm NL})$ from $\simeq 25$ to $\simeq 2$ for $l_{\rm cut}=30$. This reflects both the larger volume, since at fixed $l_{\rm min}$ the larger comoving distance of these galaxies gives access to smaller $k_{\rm min}$, and the bias model: $b_g^{\rm fid}-1$ grows from about $0.6$ at $z=1$ to about $2$ at $z=3$, so response to $f_{\rm NL}$ grows with redshift.  Conversely, $\sigma(f_{\rm NL})\simeq 25$ with $z<1$ is worse than our forecast for DESI LRGs with SO ($\simeq 14$, App.~\ref{desi}), despite twice the sky area and a far higher number density: LSST galaxies at $z<1$ have $b_g^{\rm fid}-1\lesssim 0.6$, so they respond weakly to $f_{\rm NL}$, and they have broader photometric redshifts, $\sigma_z/(1+z)=0.05$ against $0.027$. Second, replacing ACT with SO-like noise lowers the reconstruction noise; in our DESI-LRG forecast with $vv$ this reduces the cross-correlation-only $\sigma(f_{\rm NL})$ by 35--40\%, from 21--24 to about 14 (App.~\ref{desi}). Third, the LSST--SO overlap ($f_{\rm sky}=0.4$) is three to five times the area used by current ACT$\times$DESI kSZ analyses~\cite{Hotinli:2025tul,Chaussidon:2026vmn}.

\subsection{Galaxy $\times$ CMB lensing}\label{lensing}

CMB lensing \cite{Smith:2007rg, Lewis:2006fu} is another important CMB secondary effect, separately well measured at SNR $\sim40-50$ by ACT~\cite{ACT:2023dou}, \textit{Planck}~\cite{Planck:2018lbu}, and SPT~\cite{SPT-3G:2026uol}. The most precise measurement to date comes from a unified ACT+\textit{Planck}+SPT (APS) lensing measurement at SNR $=61$ \cite{ACT:2025qjh}. With SO, SNR is expected to rise to $\mathcal{O}(10^2)$~\cite{SimonsObservatory:2018koc}, enhancing its capability as a precision probe of cosmology. As with~$v$, lensing convergence $\kappa$ is an unbiased $\delta_m$ tracer correlated with $\delta_g$ and thus its cross correlation with galaxies constrains~$f_{\rm NL}$~\cite{2026JCAP...05..098C,Schmittfull:2017ffw,DESI:2023duv,Fabbian:2025fdk,2026A&A...710A.360B}. Measured constraints from quasar$\times$CMB-lensing cross-correlations are $f_{\rm NL}=-26^{+45}_{-40}$~\cite{DESI:2023duv} and $2^{+28}_{-34}$~\cite{2026JCAP...05..098C} for $p=1.6$, and $-13.8^{+26.7}_{-25.0}$~\cite{Fabbian:2025fdk} and $-16^{+22}_{-23}$ ~\cite{2026A&A...710A.360B} for $p=1$. We note the first of these used only cross-correlations, due to the quasar auto-spectrum showing excess large-scale power. CMB lensing correlations with the cosmic infrared background have also been used to constrain $f_{\mathrm{NL}}$ to $f_{\mathrm{NL}}=43\pm23$~\cite{2026arXiv260503783T,2023PhRvD.108h3522M}, assuming $p=1$.

{We obtain the angular spectra} via Eq.~\eqref{proj}, using window functions for $g_i$ and $\kappa$ given in Eq.~\eqref{eq_dnidz}. The $g\kappa$ cross-correlation and~$\kappa\kappa$ autospectra are computed with our modified version of \href{https://github.com/jcayuso/ReCCO}{\faGithub\ ReCCO}. The SO-like lensing reconstruction noise is computed assuming a standard-ILC MV estimator, which combines SO forecasted noise for the $TT$, $EE$, and $BB$ autospectra. The power spectra used assume SO ``goal" level sensitivity and masking $f_{\rm sky}=40\%$ (the same configuration as we use for the kSZ velocity reconstruction). Due to the wide CMB lensing projection kernel $W_{\kappa}(\chi)$, we cannot perform direct tomography with it, and so  only the galaxies are divided into $N$ redshift bins; the $g^i\kappa$ cross correlation then allows us to perform implicit tomography on $\kappa$. We then compare the $gv$ cross correlation with $g\kappa$ for constraining~$f_{\rm NL}$.
\begin{figure}
    \centering
    \includegraphics[width=\linewidth]{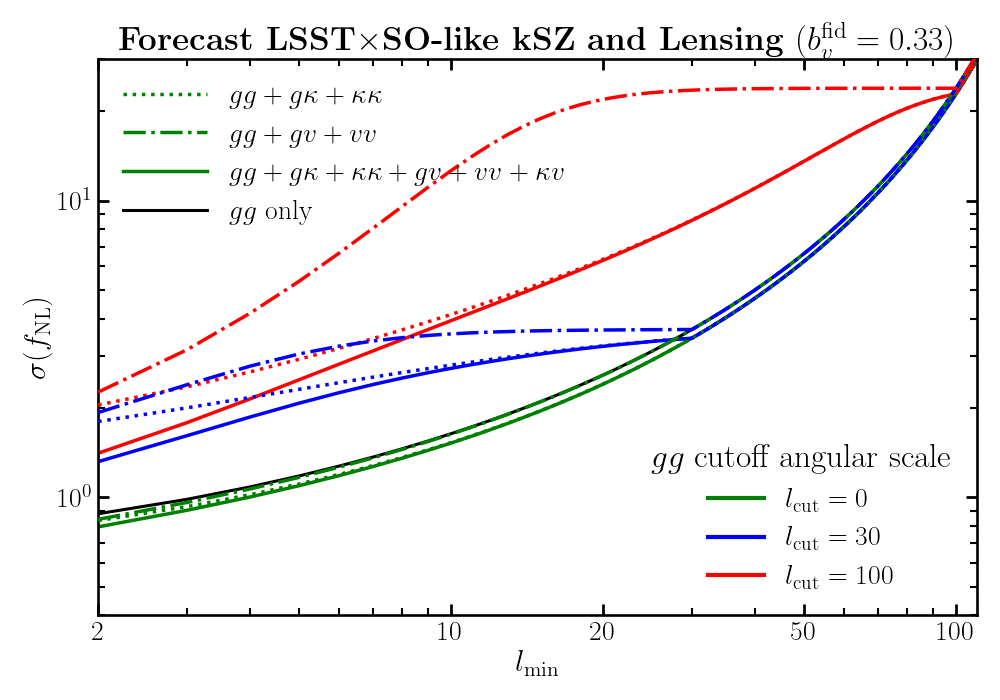}
    \caption{LSST $\times$ CMB lensing compared with LSST $\times$ kSZ when $b_v^{\rm fid}=0.33$, both reconstructed assuming SO-like noise levels. The forecast with $g\kappa$ is competitive with $gv$ for all values of $l_{\rm cut}$. In particular, $g\kappa$ forecasts tighter constraints at $l\lesssim l_{\rm cut}$, but $gv$ catches up as both approach $l_{\rm min}$, where the velocity signal begins to dominate over noise. The \textit{all case} {$gg(l>30)+g\kappa+\kappa\kappa+gv+vv+\kappa v$} reaches $\sigma(f_{\rm NL})\sim 1.35$. There is noticeable improvement over the kSZ \textit{baseline} of $\sigma(f_{\rm NL})\sim 1.95$.}
    \label{gkappa_lsst_bv=0.33}
\end{figure}

We show in Fig.~\ref{gkappa_lsst_bv=0.33} the forecast for LSST. We see that $g\kappa$ initially constrains $f_{\rm NL}$ better than $gv$ at slightly higher $l_{\rm min}$, but $g\kappa$ approaches the constraint obtained by $gv$ at the largest scales. Physically, this is because $\kappa$ traces $\delta_m$ independently of scale $k$, while the continuity equation relates radial velocity to density $\delta_m$, $v(k)=i\mu (faH)\delta_m/k$ at linear level, where $\mu=\hat{k}\cdot\hat{n}$. The $1/k$ dependence thus amplifies the velocity signal at low-$k$ (large scales) and suppresses it at higher-$k$ (smaller scales), relative to the noise. 

\begin{figure}
    \centering
    \includegraphics[width=\linewidth]{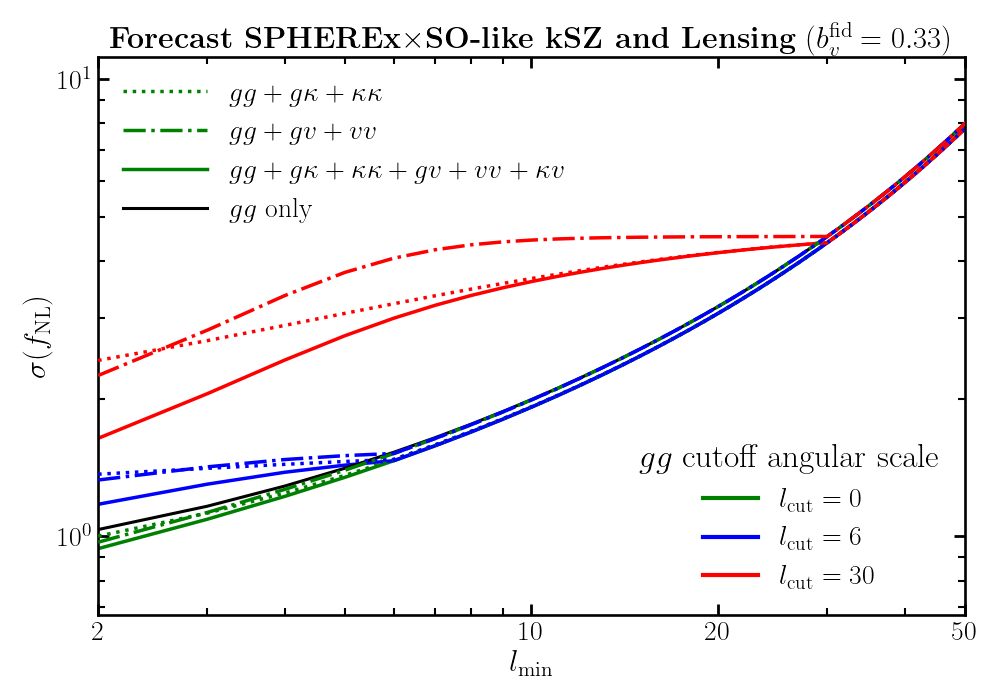}
    \caption{Forecasted SPHEREx $g\kappa$ cross correlation only constraint, compared with $gv$ at $b_v^{\rm fid}=0.33$. Interestingly, cross correlating CMB lensing with galaxies constrains $f_{\rm NL}$ more tightly than kSZ at almost all scales except the largest $l_{\rm min}=2$. When combining $g\kappa+gv$ cross correlations, we forecast $\sigma(f_{\rm NL})$ constraints tightening to $1.15$ for cross correlations alone.}
    \label{fig:spherex_lensing}
\end{figure}

SPHEREx has less redshift overlap with $\kappa$ compared to LSST --- the broad CMB lensing kernel peaks around $z\simeq 2$, while the SPHEREx number density is dominant within the low-redshift regime with $dn/dz$ peaking at $z<1$.  However, given the low intrinsic kSZ signal amplitude $b_v^{\rm fid}=0.33$, the lensing cross correlation is forecasted to be more constraining than kSZ for every $l_{\rm min}\geq 3$ (as shown in Fig.~\ref{fig:spherex_lensing}), with $\sigma(f_{\rm NL})$ at $l_{\rm min}=2$ being only slightly larger.

\subsection{Combining Galaxy $\times$ kSZ Velocity and Galaxy $\times$ CMB lensing}

Since $f_{\rm NL}$ constraints from galaxies cross-correlated with $\kappa$ are competitive with $b_v^{\rm fid}=0.33$ kSZ (Figs.~\ref{gkappa_lsst_bv=0.33} and~\ref{fig:spherex_lensing}), a natural extension is to consider cross-correlations with all three tracers $gg(l>l_{\rm cut})+gv+vv+g\kappa+\kappa\kappa+\kappa v$. We show in Figs.~\ref{gkappa_lsst_bv=0.33} and~\ref{fig:spherex_lensing} the combined constraining power of $gv+g\kappa$ cross correlations. We observe that $\sigma(f_{\rm NL})$ is forecasted to improve from this combination, reaching $\sigma(f_{\rm NL})\sim 1.35$ ($\sim 1.15$) assuming LSST (SPHEREx) $b_v^{\rm fid}=0.33$.

\begin{figure}
    \centering
    \includegraphics[width=\linewidth]{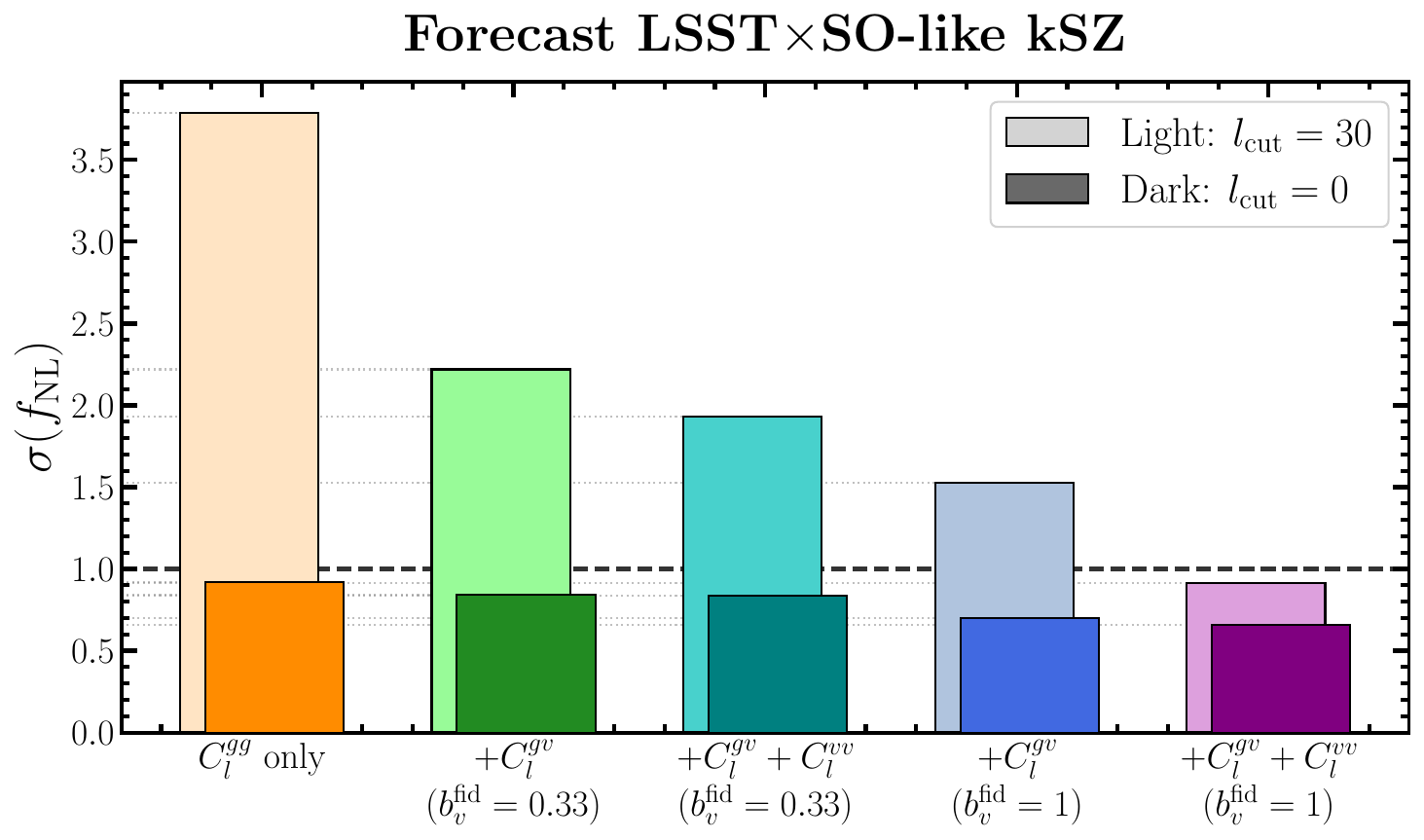}
    \caption{Forecast for LSST galaxies cross correlated with kSZ velocity reconstructed from SO+LSST.
    We emphasize that $\sigma(f_{\rm NL})<1$ is obtained from $gg(l>30)+gv+vv$ when $b_v^{\rm fid}=1$ (light-purple), while $b_v^{\rm fid}=0.33$ cross correlations alone (teal) reduces $\sigma(f_{\rm NL})$ by almost half compared to galaxies only.}
    \label{barchart_lsst}
\end{figure}

\begin{figure}
    \centering
    \includegraphics[width=\linewidth]{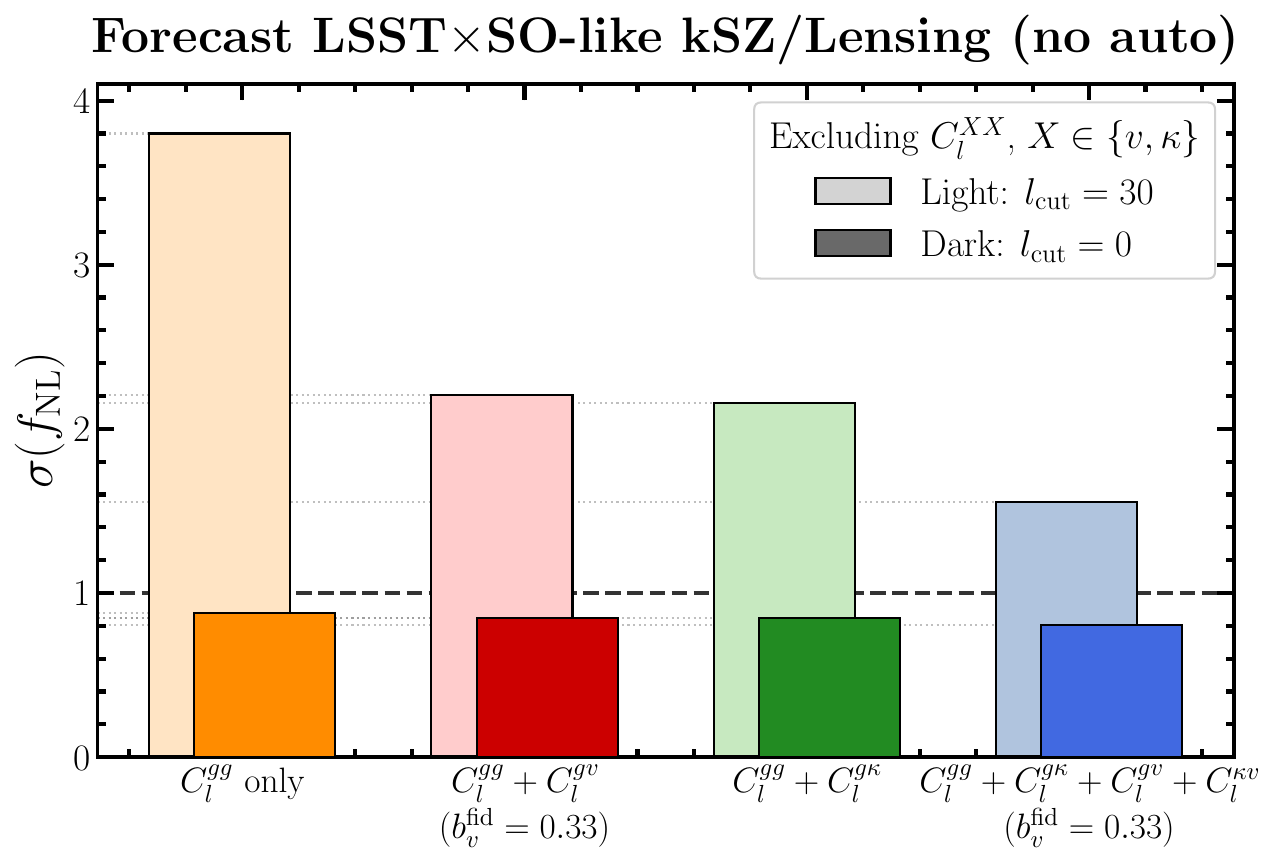}
    \caption{Forecast for LSST galaxies cross correlated with SO lensing/SO+LSST reconstructed kSZ velocity with $b_v^{\rm fid}=0.33$, without the $vv$ and $\kappa\kappa$ autospectra. The $g\kappa$ and $gv$ cross correlations alone separately constrain $\sigma(f_{\rm NL})\sim 2.2$. Extra variance cancellation from combining $gv+g\kappa$ tightens constraint to $\sigma(f_{\rm NL})\sim 1.5$.}
    \label{barchart_lsst_lensing_no_auto}
\end{figure}

\begin{figure}
    \centering
    \includegraphics[width=\linewidth]{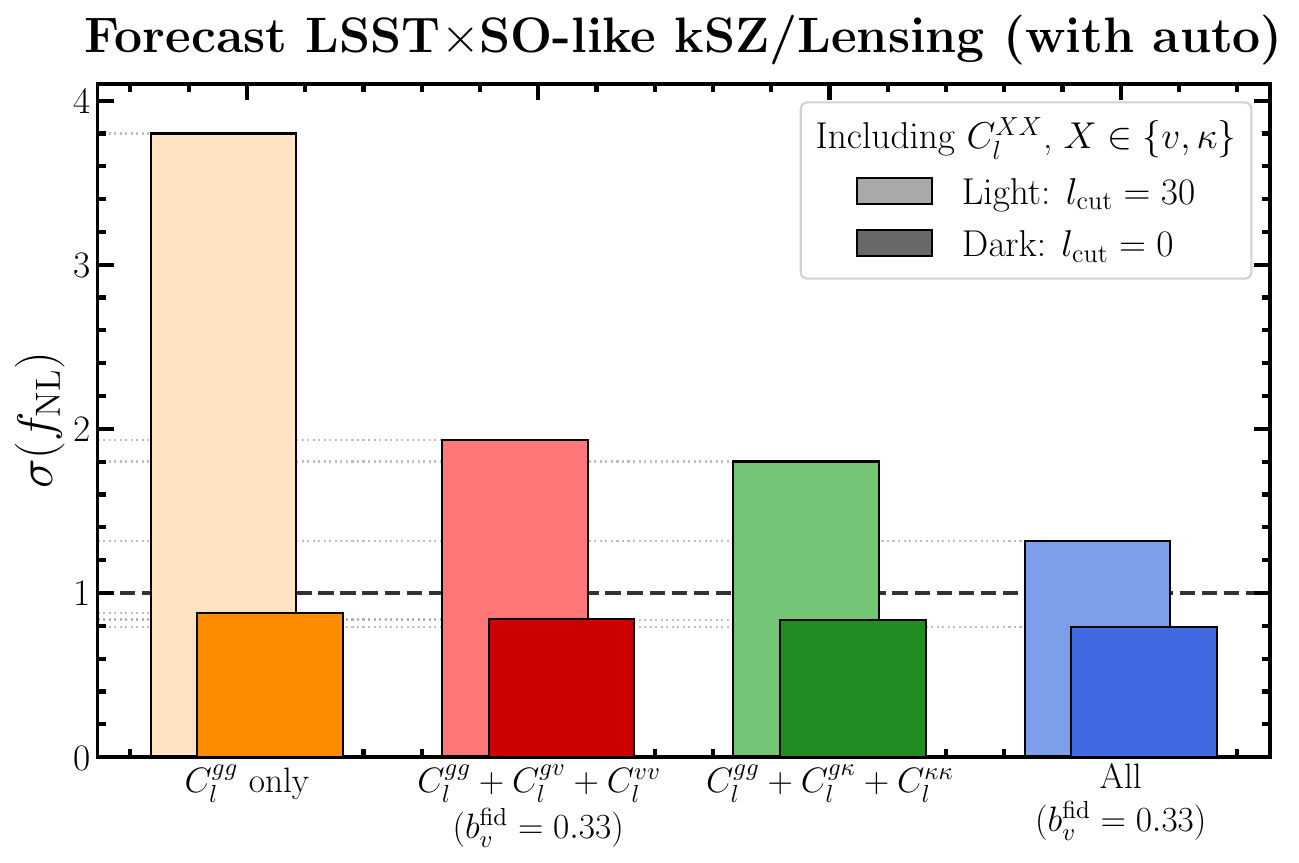}
    \caption{As in previous figure, forecast for LSST galaxies cross correlated with SO lensing/SO+LSST reconstructed kSZ velocity with $b_v^{\rm fid}=0.33$, but including the velocity and lensing autospectrum. CVC slightly improves separate $gv$ and $g\kappa$ cross correlations alone constraints to $\sigma(f_{\rm NL})<2$, compared to Fig.~\ref{barchart_lsst_lensing_no_auto}. We note $\sigma(f_{\rm NL})\sim 1.3$ is obtained from combining $gv+g\kappa$ cross correlations alone.}
    \label{barchart_lsst_lensing}
\end{figure}

\section{Results and Discussion}\label{result}

\begin{figure}
    \centering
    \includegraphics[width=\linewidth]{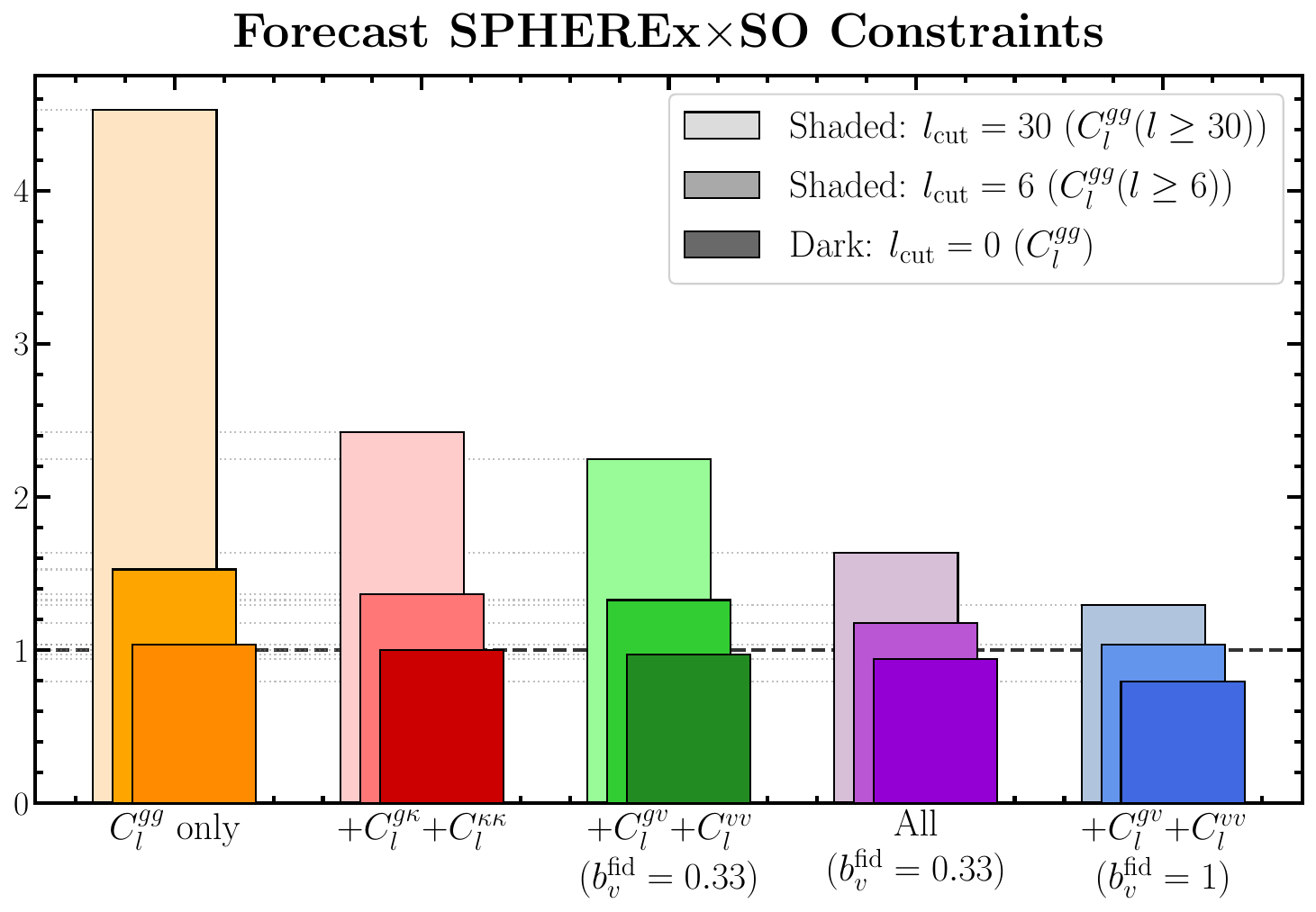}
    \caption{Forecast for SPHEREx galaxies cross correlated with kSZ velocity reconstructed from SO+SPHEREx. With $l_{\rm cut}=6$, Baseline $b_v^{\rm fid}=0.33$ $gv$ and $g\kappa$ cross correlations alone both constrain $\sigma(f_{\rm NL})\sim 1.3$, with optimistic $b_v^{\rm fid}=1$ giving $\sigma(f_{\rm NL})\sim 1$. Their combination reaches $\sigma(f_{\rm NL})\sim 1.15$. {Under the $b_v^{\rm fid}=0.33$ baseline, CVC from the $vv$ and $\kappa\kappa$ autospectra (as well as their combination) is insignificant.}}
    \label{barchart_spherex}
\end{figure}

We summarize our forecasted $\delta_g\times v$ cross correlation constraints on $f_{\rm NL}$ in bar charts, Figs.~\ref{barchart_lsst} - \ref{barchart_spherex}. Darker shades {assume} full-scale $gg$ information, shown for reference purposes. We have shown that with $b_v^{\rm fid}=1$ cross correlations alone, we reach the $\sigma(f_{\rm NL})\sim 1$ threshold for both LSST and SPHEREx $\times$ SO. Despite lower intrinsic amplitude of the velocity signal ($b_v^{\rm fid}=0.33$), {we forecast appreciably tighter constraints for both galaxy surveys (Fig.~\ref{barchart_lsst} and~\ref{barchart_spherex}, light green)} on $f_{\rm NL}$ with $gv$ cross correlations alone. For LSST, we also find constraints from $g\kappa$ competitive with $gv$ when $b_v^{\rm fid}=0.33$ (Fig.~\ref{barchart_lsst} for kSZ velocity, Fig.~\ref{barchart_lsst_lensing} for lensing, both light green). This motivates a simultaneous $g\kappa$ and $gv$ cross correlation in our Fisher forecast. We report a sizable $\sigma(f_{\rm NL})$ improvement over including either $gv$ or $g\kappa$ alone (Fig.~\ref{barchart_lsst_lensing}, blue), reduced to $\sigma(f_{\rm NL})<1.35$ with $gv+g\kappa$ cross correlations alone ($b_v^{\rm fid}=0.33$).\footnote{We note our forecasted results improve by $\sim 20\%$ when assuming a different LSST galaxy bias $b_g^{\rm fid}(z)=1+z$ (following~\cite{Schmittfull:2017ffw}), which we used in an earlier version of the forecast.} For SPHEREx, the maximum scale to which SPHEREx effectively mitigates galaxy clustering systematics will also be crucial (Fig.~\ref{barchart_spherex}).

Earlier forecasts give $\sigma(f_{\rm NL})=1.8$ ($1.2$) for SO lensing with the LSST gold (optimistic) sample when keeping the galaxy autospectrum down to $l_{\rm min}=20$, and $1$ for kSZ tomography with SO and LSST~\cite{SimonsObservatory:2018koc}, the latter of which remains the SO target~\cite{SimonsObservatory:2025wwn}. For LSST cross correlated with a CMB-S4-like survey, kSZ tomography~\cite{Munchmeyer:2018eey} forecasts $0.45$, close to our $gg+gv+vv$ results of $0.66$ ($0.84$) for $b_v^{\rm fid}=1$ ($0.33$). For CMB-S4 lensing with LSST at $f_{\rm sky}=0.5$, Ref.~\cite{Schmittfull:2017ffw} finds that discarding $gg$ at $l\le 18$, while keeping $g\kappa$ and $\kappa\kappa$ down to $l_{\rm min}=2$, raises $\sigma(f_{\rm NL})$ from $0.4$ to $0.7$ (their Fig.~18). Our lensing forecast ($gg+g\kappa+\kappa\kappa$) goes from $0.83$ to $1.80$ when $gg(l<30)$ is discarded. To our knowledge, these are the first forecasts for future surveys that focus on $f_{\mathrm{NL}}$ constraints from the cross spectra of external tracers with the galaxy distribution alone, as opposed to CVC improvements.

\section{Conclusion}\label{concl}

In this paper, we have focused on using 2D redshift tomography to forecast constraining power on local PNG from $\delta_g\times v$ and $\delta_g\times\kappa$ cross correlations alone. \textit{Baseline scenarios} with LSST and SPHEREx inclusive of $vv$ forecast $\sigma(f_{\rm NL})\sim 0.9-1.9$ and $\sim 1.0-1.35$, respectively. Without $vv$, \textit{baseline} constraints degrade to $\sigma(f_{\rm NL})\sim 1.5-2.2$ and~$\sim 1.35-1.5$. The higher end of the reported error bar ranges is due to $b_v^{\rm fid}=0.33$ weakening the $gv$ cross correlation signal. {We forecast that $g\kappa$-alone gives similar constraints as $gv$-alone when $b_v^{\rm fid}=0.33$,} but we emphasize that with all three tracers $\delta_g,v,\kappa$, cross correlations alone {including $vv$ and $\kappa\kappa$ autospectra} is forecasted to improve $\sigma(f_{\rm NL})$ to $\sim 1.3$ $ (\sim 1.15)$.  Thus, measuring values of this bias at higher redshifts and calibration of the optical depth are required to more definitively quantify their impact on this cross correlation. There are additional factors to account for: in both surveys, the error bar can potentially inflate due to large-scale $gg$ systematic residuals in the covariance, not considered in these forecasts. Another caveat arises from the absence of foreground considerations in our velocity reconstruction, but which is not a major concern due to $gv$ foreground mitigation techniques that do not deprecate the SNR~\cite{EmbilVillagra:2026fnh}. We conclude that relying on cross-correlations alone and redshift tomography, the next generation of CMB (SO) $\times$ LSS (LSST, SPHEREx) surveys has promising potential for narrowing down $\sigma(f_{\rm NL})$ to approach order unity, given systematics hard to mitigate from the large-scale galaxy autospectrum.

\begin{acknowledgments}
We thank Olivier Doré, Simone Ferraro, Boryana Hadzhiyska, and Elisabeth Krause for helpful discussions. We also especially appreciate Elisabeth Krause for detailed comments on an early draft that improved this work. G.L. is supported by the Cambridge International Isaac Newton Scholarship. FMcC is supported by a Royal Society University Research Fellowship. BDS is supported by the European Union (ERC, CosMicBacklight, 101229997). Support for this work was provided by NASA through the NASA Hubble
Fellowship grant HST-HF2-51616.001 awarded by the Space Telescope
Science Institute, which is operated by the Association of Universities for
Research in Astronomy, Incorporated, under NASA contract NAS5-
26555. This is not an official SO Collaboration paper. 
\end{acknowledgments}

% \clearpage
\appendix

\section{Fisher Analysis}\label{fisher}
The bandpower Fisher formalism is particularly suitable for our cross-correlations-alone analysis as it allows us to easily drop the auto spectra while retaining the cross. Given the noise-inclusive total power spectra, the bandpower Fisher matrix for a vector of parameters $\Pi$ is:
\begin{equation}
    \mathcal{F}_{\Pi_i \Pi_j} = \sum_{l=l_{\rm min}}^{l_{\rm max}}\frac{\partial \hat C_l^T}{\partial\Pi_i }\mathcal{C}_l^{-1}\frac{\partial \hat C_l}{\partial\Pi_j}
\end{equation}
where $l_{\rm max}=200$, the cutoff scale below which we assume linear bias on $\delta_g$  is a good approximation, {and $l_{\rm min}$ is the largest-scale multipole accessible in our power spectra.  The covariance matrix $\mathcal{C}$ is given by} 
\begin{equation}
    \mathcal{C}(\hat C_l^{\alpha\beta}, \hat C_l^{\gamma\delta}) = \frac{1}{(2l+1)f_{\rm sky}}(\hat C_l^{\alpha\gamma}\hat C_l^{\beta\delta}+\hat C_l^{\alpha\delta}\hat C_l^{\beta\gamma}),
\end{equation}
{where $f_{\rm sky}$ is the observed fraction of the sky after masking}. {The datavector $\hat C_l$ is formed by all elements in the {noise-inclusive} upper-triangular matrix of
$\begin{bmatrix}
     C_l^{gg}+N_l^{gg} & C_l^{gv} \\ C_l^{vg} & C_l^{vv} + N_l^{vv}
\end{bmatrix}$}. With $N$ redshift bins, there is a set of $2N$ bias parameters $\{b_{g,i},\cdots b_{v,i}\}$ which are marginalized over, and the parameter of interest $f_{\rm NL}$. Thus, the Fisher parameter set $\Pi$ consists of $\Pi=\{b_{g,i},b_{v,i}, f_{\rm NL}\}_{i\in [1,N]}$. From this we calculate the Cramer-Rao lower bound on $\sigma(f_{\rm NL})=\sqrt{(\mathcal{F}^{-1})_{f_{\rm NL}f_{\rm NL}}}$. For SPHEREx, where $gg$ covers more sky than the cross-spectra, we add the Fisher matrices of two patches.

\section{kSZ Velocity Estimator}\label{ksz_estim}

{In the squeezed limit where we reconstruct large-scale kSZ velocity, we} can derive the following estimator for $v$ in harmonic space:

\begin{equation}
    \hat{v}_{LM}^i = A_L^i\sum_{l m; l' m'}(-1)^M \begin{pmatrix} l & l' & L \\ m & m' & -M \end{pmatrix} \omega^i_{l l' L}T_{l m}\delta ^{g^i}_{l' m'},
\end{equation}
where the squeezed limit demands that $L\ll l, l'$,  $\omega^i_{ll'L}$ is the {optimal} weighting to be determined, $A^i_L$ is a normalization to ensure unbiasedness, and $\begin{pmatrix} l & l' & L \\ m & m' & -M \end{pmatrix}$ indicates the Wigner-3J symbols. This ensures rotational invariance of the $\langle\delta_gT_{\rm kSZ}\rangle\sim \langle\delta_g\dot{\tau}v\rangle$ bispectrum (where $\dot{\tau}=d\tau/d\chi$), while $T_{lm}$ and $\delta^{g^i}_{l'm'}$ are angular multipole moments obtained by projections into harmonic space:
\begin{equation}
    X_{lm} = \int_{\Omega} d^2\hat{n}Y^\ast_{lm}(\hat{n})X(\hat{n}),
\end{equation}
with $Y_{lm}(\hat n)$ the spherical harmonics on the sphere, and the star indicating their complex conjugate. Using the squeezed limit galaxy-kSZ cross correlation,
\begin{align*}
\left\langle T_{l m}^{\mathrm{kSZ}} \delta_{l'm'}^{g^i} \right\rangle = \sum_{LM}(-1)^M \begin{pmatrix} l & l' & L \\ m & m' & -M \end{pmatrix} f_{l l' L}  C_{l'}^{\tau g^i} v^i_{LM},
\end{align*}
where
\begin{equation*}
    f_{l l' L} = \sqrt{\frac{(2l + 1)(2l' + 1)(2L + 1)}{4\pi}} \begin{pmatrix} l & l' & L \\ 0 & 0 & 0 \end{pmatrix},
\end{equation*}
{and $C_l^{g_i\tau}$ is the galaxy-optical depth correlation, which we model  with a halo model prescription assuming Battaglia-like profiles \cite{Battaglia:2016xbi} for the electrons} (see Appendix~A of \cite{McCarthy:2024nik}). Any mismodelling of $C_l^{g_i\tau}$ (whether from using the halo model or other models) is absorbed into the velocity bias $b_v$. Demanding that the estimator is an unbiased estimator of the true velocity $v_{r LM}^i$, we obtain
\begin{equation}
    A^i_L\sum_{l;l'}\omega^i_{ll'L}f_{ll'L}C_{l'}^{\tau g_i}=2L+1.
\end{equation}
Under this constraint, we minimize the variance of $\hat{v}^i_{r LM}$ with a Lagrange multiplier. This gives
\begin{align}
    \omega^i_{ll'L} &= \frac{f_{ll'L}C_{l'}^{\tau g_i}}{\hat C_{l}^{TT}\hat C_{l'}^{g_ig_i}};\\
    A_L^{i} &= (2L + 1) \left( \sum_{l; l'} \frac{\left( f_{l l' L} C_{l'}^{g_i\tau} \right)^2}{\hat C_l^{TT} \hat C_{l'}^{g_ig_i}} \right)^{-1}.
\end{align}

\section{Estimation of Signal to Noise}\label{snr_estimate}
In this appendix, we discuss the effect of the velocity bias $b_v$ on the SNR of $gv$ and $vv$. We start with the noise-inclusive total $gg$, $gv$, and $vv$ power spectra $A(b_g^2C_l^{gg})+N_l^{gg}, B(b_gb_vC_l^{gv}),$ and $C(b_v^2C_l^{vv}) + N_l^{vv}$, where $A$, $B$, and $C$ are amplitudes. The overall signal-to-noise ratio (SNR) for $gv$ is given by 
$\sqrt{\sum_{l} \left. F^{BB}_l\right|_{B=1}}$. Here, the Fisher matrix contributed by mode $l$ is $F_l=(C_l^{gv})^T(\mathcal{C}_l^{gv})^{-1}C_l^{gv}|_{A=B=C=1}$ and $\mathcal{C}_l$ is the bandpower covariance of the noise-inclusive signal defined above. Writing in schematic matrix notation and ignoring bin indices,
\begin{equation*}
    F_l^{gv} \sim (C_l^{gv})^2\left[(C_l^{gg}+N_l^{gg}b_g^{-2})(C_l^{vv}+N_l^{vv}b_v^{-2})+(C_l^{gv})^2\right]^{-1}.
\end{equation*}
An entirely analogous argument gives for $C_l^{vv}$:
\begin{equation}
    F_l^{vv} \sim (C_l^{vv})^2\left[C_l^{vv}+N_l^{vv}b_v^{-2}\right]^{-2}.
\end{equation}
This completes our argument that the impact of $b_v$ on cross correlations alone is suppressed when the velocity autospectrum signal dominates over reconstruction noise.

\section{Binning and Correlation across Redshift Bins}\label{binning}
Idealistically we can cross-correlate every field $X_i$ and $Y_j$ ($1\leq i,j \leq N_{\rm bins}$) and bin the redshift space as finely as possible, but this is computationally challenging when inverting heavy covariance matrices and will present a large analysis challenge with many elements in a datavector. Through this forecast, we investigate the optimal LSST/SPHEREx binning and cross-correlation strategy, i.e. the number of bins and the furthest redshift bin correlations we need to saturate $\sigma(f_{\rm NL})$ constraining power.

For LSST, we have tested that constraints are saturated at $N_{\rm bins}=18$ by reducing $N_{\rm bins}$ incrementally from there, but always such that the bins are defined by having equal width $\Delta\chi$ in comoving distance.

The next step is to vary the number of $gv$ off-diagonals of the $C_l^{gv}$ matrix block. An off-diagonal correlation, e.g. $C_l^{g_iv_j}$, correlates $\delta_g$ in the $i^{th}$ redshift bin with $v$ in the $j^{th}$ bin. Thus when we increase the number of off-diagonals, we are correlating $\delta_g$ and $v$ from redshift bins farther away, which become less and less correlated. {Theoretically, there is very little} information contained in $C_l^{g_iv_i}$ because $\delta_g$ is a scalar field while $v$ is a radial vector field: it is equally likely that matter traced by $v$ within the $i^{th}$ bin is moving away ($v>0$) or moving towards us ($v<0$) due to statistical isotropy, thus cancelling out each other~\cite{EmbilVillagra:2026fnh} (however, one should note that due to large-scale velocity field evolution along the kSZ lightcone over the redshift range of each bin, there is a small amount of information in $C_l^{g_iv_i}$). With inter-bin correlations, $gv$ will form a nonzero radial dipole because gravity from galaxies in the $i^{th}$ bin will pull matter traced by $v$ within the $j^{th}$ bin towards the $i^{th}$ bin. 

\begin{figure}
    \centering
    \includegraphics[width=\linewidth]{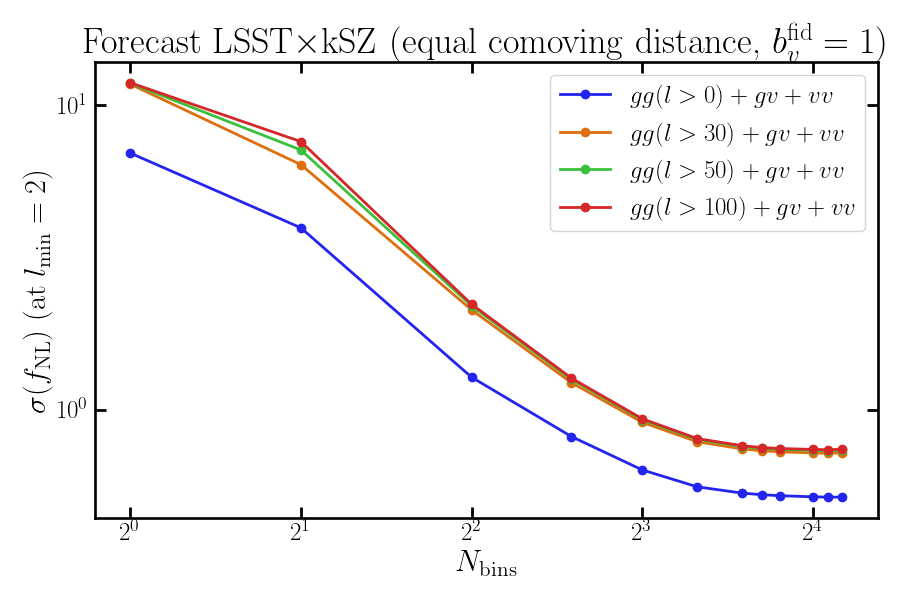}
    \caption{The dependence of $\sigma(f_{\rm NL})$ on $N_{\rm bins}$. Constraints are saturated at $N_{\rm bins}=12$, which equipartitions $\chi-$space bounded by $z=0.02$ and $z=4$ into $11$ sections.}
    \label{Nbins_exp}
\end{figure}

\begin{figure}
    \centering
    \includegraphics[width=\linewidth]{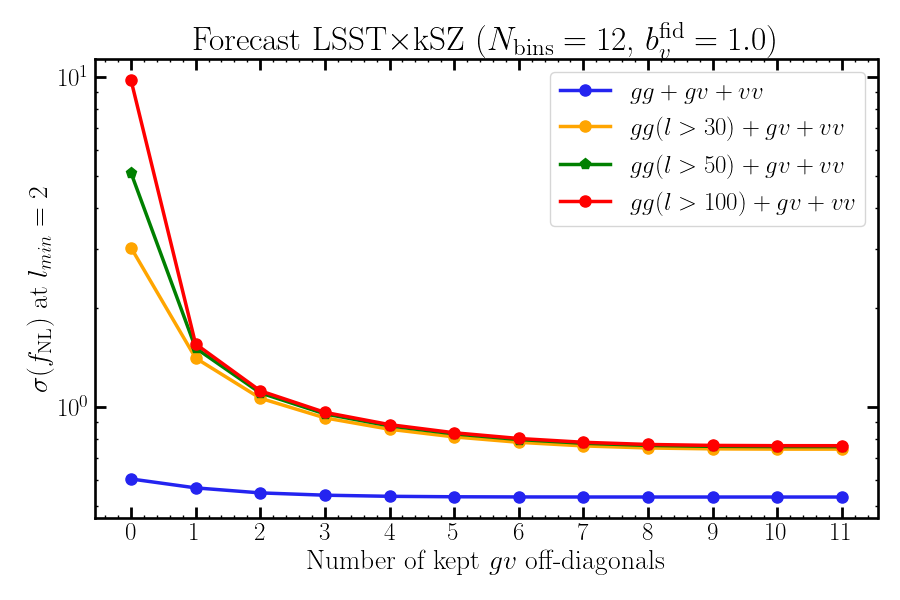}
    \caption{The dependence of $\sigma(f_{\rm NL})$ on the number of kept $C_l^{gv}$ off-diagonals (keeping all of $gg+vv$). Zero off-diagonal means only correlation between $\delta_g$ and $v$ from the same $z-$bin is accounted for. As expected, $gv$-diagonal alone contains no information (modulo lightcone evolution effects) and minimally affects $\sigma(f_{\rm NL})$. The constraint effectively saturates beyond off-diagonals $=3$ for $l_{\rm cut}=0$ and off-diagonals $=6$ for $l_{\rm cut}=30, 50, 100$.}
    \label{fig:gv off-diags}
\end{figure}

From Figs.~\ref{Nbins_exp} and~\ref{fig:gv off-diags}, we observe that accounting for correlations between the six-nearest bins saturates the $\sigma(f_{\rm NL})$ constraint for $N_{\rm bins}=12$, which equipartitions $0.02<z<4$ into eleven sections. Instead of cross-correlating every field in every redshift bin ($18\times 18=324$ correlations), we at most need $101$ correlations to reach equally good constraints, alleviating some computational challenges of the heavy covariance matrix. {In our forecasts, we used $N_{\rm bins}=18$ with all off-diagonals.} 

For the full SPHEREx $\times$ kSZ forecast, multitracer requires marginalization over $44\times 2+33+22\times 2=165$ galaxy and velocity bias parameters $b_g^i$ and $b_v^i$. Thus, the full datavector has dimensions of $331\times 332/2=54946$. We find that on a scale normalized against $N_{\rm bins}=11$, accounting for all correlations separated less than $6/22$ from each other saturates $\sigma(f_{\rm NL})$, reducing the covariance dimension down to $24967$ and yielding a $6\times$ computational speedup compared to if we account for every inter-bin correlation.

\section{DESI LRGs$\times$ACT/SO-like Forecast}\label{desi}

We show preliminary $\sigma(f_{\rm NL})$ forecasts for DESI$\times$ ACT upgraded to DESI$\times$ SO-like, focusing on photometric DESI Luminous Red Galaxies (LRGs, and their number densities) already measured now. Though simplistic, the forecasted $\sigma(f_{\rm NL})$ from this setup is similar to that obtained from DESI LRGs $\times$ ACT analysis \cite{Chaussidon:2026vmn}, and illustrates the additional constraining power from cross correlating galaxies with a lower-noise CMB survey.

\begin{figure}
    \centering
    \includegraphics[width=\linewidth]{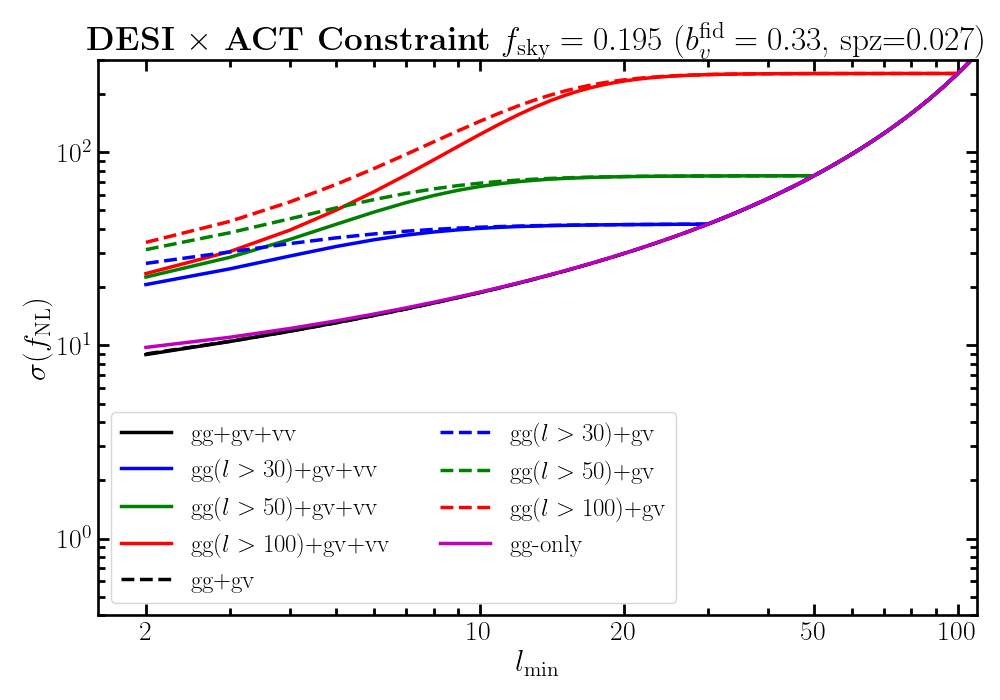}
    \caption{Using the setup described in the main text, we make a simplistic $gg(l>l_{\rm cut})+gv+vv$ cross-correlation alone forecast for DESI DRI/II LRGs $\times$ ACT. We assume $f_{\rm sky}=0.195$, photo-redshift errors of $0.027$ appropriate for LRGs, and use $16$ tomographic redshift bins dividing $z=[0.4,1]$ where LRGs lie. The number density within each bin is given in \cite{McCarthy:2025brx}. For $gg$-only and $gg+gv+vv$, we reproduce $\sigma(f_{\rm NL})\sim 10$, close to that reported in \cite{Chaussidon:2024qni}. Cross correlation alone constraints $\sigma(f_{\rm NL})\sim 21-35$.} 
    \label{fig:desixact}
\end{figure}

\begin{figure}
    \centering
    \includegraphics[width=\linewidth]{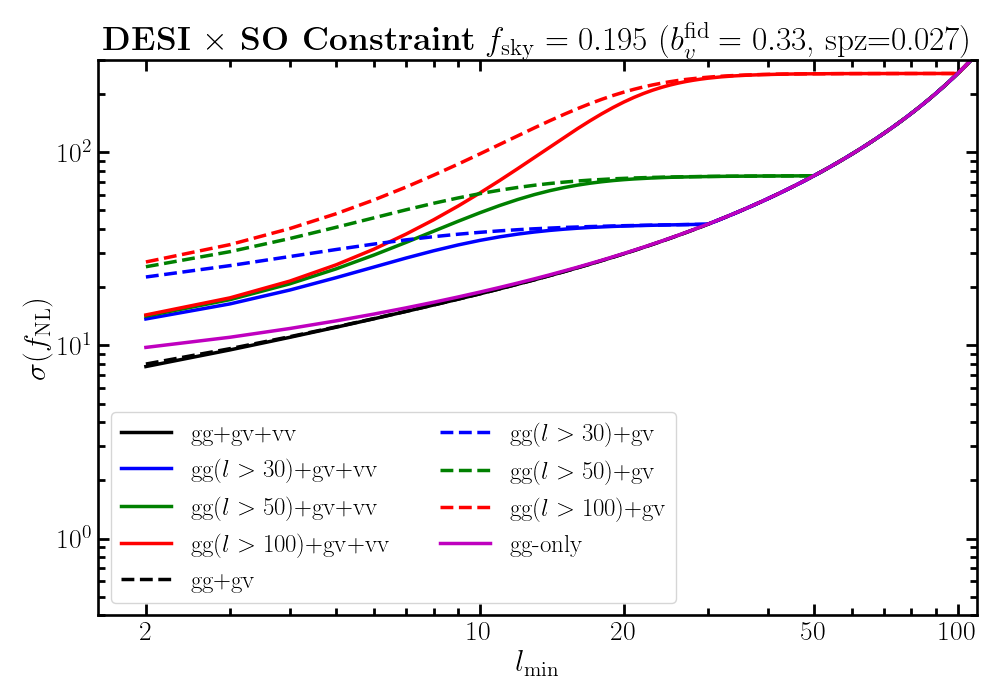}
    \caption{Under the same DESI LRG assumptions in Fig.~\ref{fig:desixact}, we make a preliminary DESI DRI/II LRGs $\times$ SO forecast, where reconstruction noise is lower with the SO-like setup. We observe that cross correlation alone constraints tighten from Fig.~\ref{fig:desixact} to $\sigma(f_{\rm NL})\sim 15$.}
    \label{fig:placeholder}
\end{figure}

\newpage
%\nocite{*}

\bibliography{main}% Produces the bibliography via BibTeX.

\end{document}